\documentclass[aps,prb,twocolumn,superscriptaddress,longbibliography]{revtex4-2}
\usepackage{amsmath,amssymb}
\usepackage[pdftex]{hyperref,graphicx}
\hypersetup{colorlinks = true, urlcolor = blue, linkcolor = blue, citecolor = blue}
\usepackage{physics}
\usepackage{xcolor}
\usepackage{bm}
\usepackage{tikz}
\usetikzlibrary{fit,arrows.meta}
\usepackage{dsfont}
\usepackage{booktabs}

\newcommand{\hc}{\text{h.c.}}
\newcommand{\ueV}{\,\mu\text{eV}}

\newcommand{\majthree}[3]{%
    \node[panel] at (#1-0.35,#2+1.55) {#3};
    \node[endm] (g1) at (#1,#2+0.75) {$\gamma_{1}$};
    \node[maj]  (g2) at (#1+2.1,#2+0.75) {$\gamma_{2}$};
    \node[maj]  (g3) at (#1+4.2,#2+0.75) {$\gamma_{3}$};
    \node[maj]  (b1) at (#1,#2-0.75) {$\bar{\gamma}_{1}$};
    \node[maj]  (b2) at (#1+2.1,#2-0.75) {$\bar{\gamma}_{2}$};
    \node[endm] (b3) at (#1+4.2,#2-0.75) {$\bar{\gamma}_{3}$};
    \node[site,fit=(g1)(b1)] {};
    \node[site,fit=(g2)(b2)] {};
    \node[site,fit=(g3)(b3)] {};
    \draw[faint,orange!90!black] (g1) -- (b1);
    \draw[faint,orange!90!black] (g2) -- (b2);
    \draw[faint,orange!90!black] (g3) -- (b3);
    \draw[faint,red!70!black]    (g1) -- (b2);
    \draw[faint,red!70!black]    (g2) -- (b3);
    \draw[faint,blue!65!black]   (b1) -- (g2);
    \draw[faint,blue!65!black]   (b2) -- (g3);
}
\begin{document}
\title{Scaling of qubit coherence in quantum dot based Majorana chains: Rabi and Ramsey oscillations of Majorana qubits formed by four and six quantum dots}
\author{Haining Pan}
\affiliation{Department of Physics, University of Florida, Gainesville, Florida 32611, USA}

\author{Sankar Das Sarma}
\affiliation{Condensed Matter Theory Center and Joint Quantum Institute, Department of Physics, University of Maryland, College Park, Maryland 20742, USA}

\begin{abstract}
We present a theoretical study of the Rabi and Ramsey coherence time of Majorana-based qubits formed by double $N$-site quantum-dot artificial Kitaev chains, following the recent experimental realization~\cite{zatelli2026majorana}.
We generalize the qubit to two chains of arbitrary length and compare the Rabi and Ramsey oscillations of the four-dot and six-dot qubits, simulated with realistic disorder for three sets of parameters: small and large superconducting gaps, and the experimental parameters.
We find that, in the pristine limit, the Rabi oscillation is independent of the chain length, whereas the Ramsey oscillation of the six-dot qubit suffers from larger leakage, because the third site suppresses the unwanted energy splitting away from the sweet spot but at the same time makes the deliberate splitting difficult to achieve, which requires a larger detuning that could excite the bulk.
In reality with disorder, neither coherence time improves universally with the chain length.
The Rabi coherence is set by the competition of three dephasing channels, and the additional dots pay off only if the fluctuation of the interchain tunneling is suppressed; 
the Ramsey coherence is set by the fluctuation of the energy splitting during the $\sigma_z$ rotation, which for the four-dot qubit is fixed by the sweet-spot disorder alone and does not depend on the mean splitting, whereas for the six-dot qubit the detuning that generates the splitting adds a fluctuation growing with the splitting itself, so the additional dots pay off only for small splittings.
This implies that a longer chain is thus a better quantum memory but not necessarily a better qubit under manipulation. 
Although topological protection is expected in a longer Kitaev chain qubit, observing it in the current Rabi and Ramsey measurement protocols remains an experimental challenge.
\end{abstract}

\maketitle

\section{Introduction}

Semiconductor nanowires in semiconductor--superconductor hybrid structures, as introduced in Refs.~\cite{sau2010generic,lutchyn2010majorana,sau2010robustness,oreg2010helical, sarma2015majorana,dassarma2023search}, are the standard topological platforms for the practical realization of non-Abelian Majorana zero modes, which can be used for topological quantum computing~\cite{kitaev2003faulttolerant,nayak2008nonabelian}. There has been impressive recent experimental progress in the semiconductor nanowire system spearheaded by the Microsoft Corporation, which adopted the semiconductor--superconductor hybrid as their quantum computing platform~\cite{microsoftquantum2023inasal,aghaee2025interferometric,aghaee2025distinct}. The current work is on the closely related alternate idea~\cite{sau2012realizing} for realizing Majorana zero modes using a 1D lattice chain of semiconductor quantum dots, which, under precisely defined conditions, may mimic the Kitaev chain idea~\cite{kitaev2001unpaired}, not literally, but qualitatively. The current work focuses on an important recent experimental breakthrough~\cite{zatelli2026majorana}, and asks the key question: How does the qubit coherence in this quantum dot Majorana platform scale with an increasing number of dots?

We recently proposed a specific qubit scheme~\cite{pan2025rabi} for the quantum dot Majorana platform using four quantum dots, configured as a double two-dot structure, which enables qubit operations as well as the possibility of Rabi and Ramsey oscillations, which are the hallmarks of coherent two-level quantum dynamics. In Ref.~\cite{zatelli2026majorana}, this qubit scheme~\cite{pan2025rabi} was implemented in an InSb-Al quantum dot hybrid platform, leading to impressive measurements of both Rabi and Ramsey oscillations and their effective coherence times, $T_{2,\text{Rabi}}$ and $T_{2,\text{Ramsey}}$. The system is akin to a simple two-dot charge qubit system containing one electron tunneling between the dots~\cite{petersson2010quantum}, except that the Majorana physics provides algebraic protection of the qubits, enhancing the coherence times compared with the unprotected standard charge qubit, where the coherence times are very short. Amazingly, the experiment~\cite{zatelli2026majorana} quantitatively verifies our predicted Rabi times~\cite{pan2025rabi} approximately.

Given this agreement between theory~\cite{pan2025rabi} and experiment~\cite{zatelli2026majorana}, the natural question that arises is whether the coherence can be further improved by adding more dots to the system, so that the coherence is long enough for braiding operations to be feasible in order to explicitly verify the non-Abelian nature of the underlying Majorana modes (although their protection is not exponential). In the current work, we theoretically generalize the problem to a double $N$-dot system, i.e., a single qubit formed by two chains of an arbitrary number ($N$) of dots each, and then explicitly calculate the coherence times for Rabi and Ramsey oscillations for $N=3$, i.e., a double three-dot system made of six quantum dots, using realistic parameter values corresponding to the experimental setup of Ref.~\cite{zatelli2026majorana}.

We find that neither $T_{2,\text{Rabi}}$ nor $T_{2,\text{Ramsey}}$ increases universally in going from $N=2$ to $N=3$; whether it increases at all depends on the relevant sample details and parameter tuning. 
Some of the relevant parameters are intrinsic, such as the induced superconducting gap (which could be large or small---``large'' being always better), but others arise from the unknown disorder affecting many aspects of the platform. 
This makes any assertive statement about the coherence increasing by some magnitude impossible, because it all depends on the parameter details in a complex manner, as shown in our results. 
Generally, $T_{2,\text{Rabi}}$ could be enhanced substantially only if the tunneling disorder is suppressed well below the values considered here, and the enhancement in $T_{2,\text{Ramsey}}$ is likewise modest at the operating splitting. 
But the important lesson is that the experimentalists may be able to fine-tune the system parameters to enhance coherence by following our detailed results. We emphasize that the Majorana zero modes in these quantum dot systems are inherently fragile because of the very short length of the system, and as such they belong to what was called ``ugly'' Majoranas in our earlier work~\cite{pan2020physical} in the context of the nanowire physics. More recently, a new terminology, ``poor man's Majorana,'' has been used to describe these ``ugly'' Majoranas in quantum dot systems~\cite{leijnse2012parity}. The key conceptual point already emphasized well in the original paper~\cite{sau2010generic} is that these are not Majoranas in the classic sense of exponential protection~\cite{kitaev2001unpaired}; these are protected algebraically---these should be thought of as high-quality ``charge qubits''~\cite{petersson2010quantum}, where decoherence is suppressed by fine-tuning the Majorana physics.

Our work implies that one may have to do considerable system fine-tuning as well as go to larger $N$ ($N \gg 3$) for the coherence to be long enough to enable braiding operations, since it is the Ramsey coherence time, corresponding to the free-induction-decay time $T_2^*$, which determines the actual fidelity of qubit operations. By contrast, $T_{2,\text{Rabi}}$ corresponds to the Hahn echo time $T_2$, which has to be maintained by echo pulses on the qubit. We provide detailed quantitative results comparing the double two-dot and double three-dot structures with each other to establish the level of coherence protection expected with increasing $N$, and the extent to which this quantum dot Majorana platform differs from the simple semiconductor dot charge qubits, which are known to dephase quickly. We do, however, mention that the corresponding Rabi and Ramsey measurements, although already predicted theoretically~\cite{sau2025capacitancebased}, have not yet been performed in the semiconductor nanowire platform~\cite{microsoftquantum2023inasal,aghaee2025interferometric,aghaee2025distinct}. Therefore, the direct observation of Rabi and Ramsey oscillations in the Majorana quantum dot platform is certainly an important development, because such oscillations establish that these semiconductor--superconductor hybrid Majorana platforms do indeed demonstrate the canonical coherent quantum two-level dynamics defining qubits.

The rest of this paper is organized as follows.
Section~\ref{sec:model} sets up the single $N$-site chain, its sweet spot, and the splitting away from it, and reduces the double $N$-site chain to a two-level effective Hamiltonian.
Section~\ref{sec:results} gives the Rabi and Ramsey protocols in the pristine limit and the coherence times fitted from them in the presence of disorder, for $N=2$ and $N=3$.
We discuss in Sec.~\ref{sec:discussion} by identifying the dephasing channels that determine $T_{2,\text{Rabi}}$ and $T_{2,\text{Ramsey}}$, and the crossovers between them to compare with numerical results.
We conclude in Sec.~\ref{sec:conclusion}.
Appendix~\ref{app:splitting} derives the energy splitting in a single $N$-site chain.
Appendix~\ref{app:effective_model} derives the two-level effective Hamiltonian for a double $N$-site chain.
Appendix~\ref{app:rabi_analytical} solves the pristine Rabi oscillation for the double $N$-site chain.
Appendix~\ref{app:ramsey_N3} derives the three-site splitting at the Ramsey detuning and inverts it for the calibration $u^{(3)}$ of Eq.~\eqref{eq:calibration}.
Appendix~\ref{app:ramsey_analytical} solves the pristine Ramsey sequence for $N=3$.
Appendix~\ref{app:Gamma_exp} derives the junction coupling of the experimental parameter set.
Appendix~\ref{app:geff_average} shows the details of averaging a phase quadratic in the Gaussian disorder.

\section{Model}\label{sec:model}
\subsection{Single $N$-site quantum-dot chain}\label{sec:single_chain}
Starting from a chain of spin-polarized quantum dots~\cite{dvir2023realization,bordin2025enhanced,kulesh2025fluxcontrolled,sau2012realizing,leijnse2012parity,liu2023fusion}, we model each dot using a single spinless fermionic mode, with the Hamiltonian of an $N$-site chain as follows:
\begin{equation}\label{eq:Hchain}
    \mathcal{H}=\sum_{i=1}^{N}\mu_{i}n_{i}+\sum_{i=1}^{N-1}\left(t_{i}c_{i+1}^{\dagger}c_{i}+\Delta_{i}c_{i+1}c_{i}+\hc \right).
\end{equation}
Here $c_{i}$ annihilates a fermion on site $i$, $n_{i}=c_{i}^{\dagger}c_{i}$, and $\mu_{i}$ is the corresponding chemical potential.
$t_{i}$ and $\Delta_{i}$ are the elastic-cotunneling and crossed-Andreev-reflection amplitudes connecting sites $i$ and $i+1$, generated by the intervening proximitized segment, respectively~\cite{liu2022tunable}.
For simplicity, we omit the further-neighbor couplings $t_{ij}$ and $\Delta_{ij}$ for $|i-j|>1$~\cite{miles2024kitaev}, and ignore their phase difference (i.e., all $t_{i}$ and $\Delta_{i}$ are real).
In terms of the Majorana operators defined by $c_{i}^\dag=\frac{1}{2}\left( \gamma_{i}+i\bar\gamma_{i} \right)$, Eq.~\eqref{eq:Hchain} can be rewritten as
\begin{equation}\label{eq:singlechain}
\begin{split}
    \mathcal{H}=&\sum_{i=1}^{N-1}\left[-\frac{i}{2}(t_{i}-\Delta_{i})\gamma_{i}\bar\gamma_{i+1}
    +\frac{i}{2}(t_{i}+\Delta_{i})\bar\gamma_{i}\gamma_{i+1}\right]\\
    &+\sum_{i=1}^{N}\frac{\mu_{i}}{2}\left(1-i\gamma_{i}\bar\gamma_{i}\right).
\end{split}
\end{equation}

\subsubsection{Sweet spot}\label{sec:sweetspot}
In the two-site quantum dot system, the sweet spot is defined by $\abs{t_1}=\abs{\Delta_1}$ and $\mu_{1,2}=0$~\cite{leijnse2012parity}.
To generalize this to the $N$-site system, we consider the genuine $N$-site sweet spot (i.e., excluding the sweet spot of an effective chain with fewer sites embedded in the $N$-site chain~\cite{dourado2025majorana}) defined as
\begin{equation}\label{eq:sweetspot}
    \abs{t_i}=\abs{\Delta_i}, \qquad\mu_j=0,
\end{equation}
for $i = \left\{ 1, \dots N-1 \right\}$ and $j = \left\{ 1, \dots N \right\}$,
which continuously connects to the long-chain Kitaev limit in open boundary conditions with two desired Majorana zero modes localized at the end points~\cite{sau2012realizing}.
At the sweet spot Eq.~\eqref{eq:sweetspot} (assuming $\Delta_{i}>0$ without loss of generality), Eq.~\eqref{eq:singlechain} reduces to
\begin{equation}\label{eq:Hsweet}
    \mathcal{H} = i\sum_{i=1}^{N-1}\Delta_{i}\bar\gamma_{i}\gamma_{i+1}
\end{equation}
with the ground state energy being
\begin{equation}
    E_{0} = -\sum_{i=1}^{N-1} \Delta_{i},
\end{equation}
and the two degenerate ground states, of even ($e$) and odd ($o$) fermion parity, being
\begin{equation}\label{eq:eo_states}
    \begin{split}
        \ket{e} &= \frac{1}{\sqrt{2}}\left(\prod_{i=1}^{N} P^{+}_{i} + \prod_{i=1}^{N} P^{-}_{i}\right)\ket{0}, \\
        \ket{o} &= \frac{1}{\sqrt{2}}\left(\prod_{i=1}^{N} P^{+}_{i} - \prod_{i=1}^{N} P^{-}_{i}\right)\ket{0},
    \end{split}
\end{equation}
where
\begin{equation}\label{eq:P_pm}
    P^{\pm}_{i} = \frac{1}{\sqrt{2}}\left[1 \pm (-1)^{i-1} c_{i}^\dagger\right].
\end{equation}
For example, when $N=2$, we simply recover the two-site sweet spot wave function in Ref.~\cite{leijnse2012parity},
\begin{equation}\label{eq:two_site_states}
    \begin{split}
        \ket{e} &= \frac{1}{\sqrt{2}}\left(1 - c_{1}^{\dagger}c_{2}^{\dagger}\right)\ket{0}, \\
        \ket{o} &= \frac{1}{\sqrt{2}}\left(c_{1}^{\dagger} - c_{2}^{\dagger}\right)\ket{0},
    \end{split}
\end{equation}
and for $N=3$, we have
\begin{equation}\label{eq:logical_chain_states}
    \begin{split}
        \ket{e} &= \frac{1}{2}\left(1 - c_{1}^{\dagger}c_{2}^{\dagger}
        + c_{1}^{\dagger}c_{3}^{\dagger} - c_{2}^{\dagger}c_{3}^{\dagger}\right)\ket{0}, \\
        \ket{o} &= \frac{1}{2}\left(c_{1}^{\dagger} - c_{2}^{\dagger} + c_{3}^{\dagger}
        - c_{1}^{\dagger}c_{2}^{\dagger}c_{3}^{\dagger}\right)\ket{0}.
    \end{split}
\end{equation}

\subsubsection{Detuning away from the sweet spot}\label{sec:detuning}
We now consider the effect of detuning from the sweet spot to derive the splitting between the even and odd ground states, denoted by $\varepsilon\equiv E_e-E_o$ of the perturbed ground state energies.
Integrating out the $N-1$ pairs of bulk Majorana operators of Eq.~\eqref{eq:singlechain} gives, to leading order in the detuning (see Appendix~\ref{app:splitting}),
\begin{equation}\label{eq:epsilon}
    \varepsilon(N)=-\sum_{P\in \mathcal{P}_N}\frac{\prod_{i\in{P}_\mu}\mu_{i}\prod_{j\in{P}_\nu}\nu_{j}}{\prod_{j\in{P}_s}s_{j}},
\end{equation}
where $\mathcal{P}_N $ is the set of all possible paths that connect $\gamma_{1}$ to $\bar{\gamma}_{N}$ through the weak coupling bonds $-\frac{i}{2}\mu_{i}\gamma_{i}\bar{\gamma}_{i}$, $\frac{i}{2}\nu_{i}\gamma_{i}\bar{\gamma}_{i+1}$ and the strong coupling bonds $\frac{i}{2}s_{i}\bar{\gamma}_{i}\gamma_{i+1}$ of Eq.~\eqref{eq:singlechain}.
Near the sweet spot, $\nu_{i}\equiv\Delta_{i}-t_{i} \ll \mathcal{O}(\Delta)$ and $\mu_i\ll \mathcal{O}(\Delta)$ whereas $s_{i}\equiv\Delta_{i}+t_{i}\sim \mathcal{O}(\Delta_i)$.
For example, for $N=2$, the splitting is 
\begin{equation}\label{eq:epsilon_N2}
    \varepsilon(N=2)=-\nu_{1}-\frac{\mu_{1}\mu_{2}}{s_{1}},
\end{equation}
corresponding to the path $\gamma_{1}\to\bar{\gamma}_{2}$ (first term) and the path $\gamma_{1}\to\bar{\gamma}_{1}\to\gamma_{2}\to\bar{\gamma}_{2}$ (second term);
for $N=3$, the splitting is 
\begin{equation}\label{eq:epsilon_N3}
        \varepsilon(N=3)=-\frac{\nu_{1}\mu_{3}}{s_{2}}-\frac{\mu_{1}\nu_{2}}{s_{1}}-\frac{\mu_{1}\mu_{2}\mu_{3}}{s_{1}s_{2}},
\end{equation}
corresponding to the paths $\gamma_{1}\to\bar{\gamma}_{2}\to\gamma_{3}\to\bar{\gamma}_{3}$, $\gamma_{1}\to\bar{\gamma}_{1}\to\gamma_{2}\to\bar{\gamma}_{3}$, and $\gamma_{1}\to\bar{\gamma}_{1}\to\gamma_{2}\to\bar{\gamma}_{2}\to\gamma_{3}\to\bar{\gamma}_{3}$, respectively.
The three paths are enumerated explicitly in Figs.~\ref{fig:paths}(b-d).

\subsection{Double $N$-site quantum-dot chain}\label{sec:double_chain}
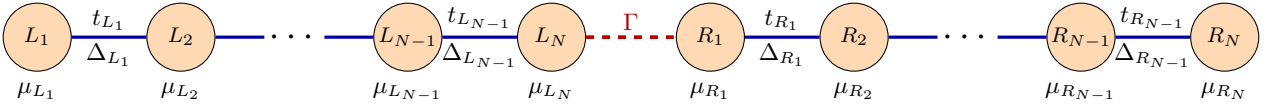
\begin{figure*}[htbp]
    \centering
    \begin{tikzpicture}[
        dot/.style={circle,draw=black,fill=orange!30,minimum size=9mm,inner sep=0pt,
            text width=8.4mm,align=center,text height=1.6ex,text depth=0.6ex,font=\footnotesize},
        bond/.style={line width=1.2pt,blue!65!black},
        tunnel/.style={line width=1.4pt,red!70!black,dashed},
        lbl/.style={font=\small,inner sep=2pt,text=black},
        ellipsis/.style={inner sep=1.5pt,font=\Large}
    ]
        \node[dot] (L1) at (0,0) {$L_1$};
        \node[dot] (L2) at (1.9,0) {$L_2$};
        \node[ellipsis] (Ldots) at (3.4,0) {$\cdots$};
        \node[dot] (LNm1) at (4.9,0) {$L_{N-1}$};
        \node[dot] (LN) at (6.8,0) {$L_N$};
        \node[dot] (R1) at (8.9,0) {$R_1$};
        \node[dot] (R2) at (10.8,0) {$R_2$};
        \node[ellipsis] (Rdots) at (12.3,0) {$\cdots$};
        \node[dot] (RNm1) at (13.8,0) {$R_{N-1}$};
        \node[dot] (RN) at (15.7,0) {$R_N$};
        \draw[bond] (L1) -- node[lbl,above] {$t_{L_1}$} node[lbl,below] {$\Delta_{L_1}$} (L2);
        \draw[bond] (L2) -- (Ldots.west);
        \draw[bond] (Ldots.east) -- (LNm1);
        \draw[bond] (LNm1) -- node[lbl,above] {$t_{L_{N-1}}$} node[lbl,below] {$\Delta_{L_{N-1}}$} (LN);
        \draw[tunnel] (LN) -- node[lbl,above,text=red!70!black] {$\Gamma$} (R1);
        \draw[bond] (R1) -- node[lbl,above] {$t_{R_1}$} node[lbl,below] {$\Delta_{R_1}$} (R2);
        \draw[bond] (R2) -- (Rdots.west);
        \draw[bond] (Rdots.east) -- (RNm1);
        \draw[bond] (RNm1) -- node[lbl,above] {$t_{R_{N-1}}$} node[lbl,below] {$\Delta_{R_{N-1}}$} (RN);
        \foreach \n/\l in {L1/L_1, L2/L_2, LNm1/L_{N-1}, LN/L_N, R1/R_1, R2/R_2, RNm1/R_{N-1}, RN/R_N}
            \node[lbl,below=2pt] at (\n.south) {$\mu_{\l}$};
    \end{tikzpicture}
    \caption{Schematic of the double $N$-site quantum-dot system.  Blue bonds denote the intrachain couplings $t_{L_i}$ and $\Delta_{L_i}$ on the left chain and $t_{R_i}$ and $\Delta_{R_i}$ on the right chain; the dashed red link denotes the interchain tunnel coupling $\Gamma$.}
    \label{fig:double_N_site}
\end{figure*}

We need two chains to construct a Majorana qubit as proposed theoretically in Ref.~\cite{pan2025rabi} and realized experimentally in Ref.~\cite{zatelli2026majorana}.
In the current work, we generalize the double two-site quantum-dot chain to the double $N$-site quantum-dot chains, as shown in Fig.~\ref{fig:double_N_site}, where two chains are connected by a gate-controlled single-electron tunnel junction.
We label the chains by $a\in\{L,R\}$ and denote site $i=1,\ldots,N$ in chain $a$ by $a_i$ (i.e., $L_i$ or $R_i$).
Therefore, the junction connects site $L_N$ of the left chain to site $R_1$ of the right chain.
The total Hamiltonian is
\begin{eqnarray}
    &&\mathcal{H}_{\text{tot}}=\mathcal{H}_L+\mathcal{H}_R+\mathcal{H}_{\text{tunn}},\label{eq:Htot}\\
    &&\mathcal{H}_{\text{tunn}}=\Gamma c_{R_1}^{\dagger}c_{L_N}+\hc,\label{eq:Htunn}
\end{eqnarray}
where $\mathcal{H}_a$ is the single-chain Hamiltonian Eq.~\eqref{eq:Hchain} of chain $a$.
Hereafter, the single-chain quantities of Sec.~\ref{sec:single_chain} evaluated for chain $a$ carry the chain label $a$, such as $c_{a_i}$, $t_{a_i}$, $\Delta_{a_i}$, $\mu_{a_i}$, $\ket{e}_a$, $E_{e,a}$, and $\varepsilon_a$.
The junction amplitude $\Gamma$, which we also take to be real, controls tunneling between the two chains.

\subsubsection{Effective model}\label{sec:effective}
Without loss of generality, we choose to work in the total even-parity subspace,
\begin{equation}
    \ket{ee}\equiv\ket {e}_L\ket {e}_R,
    \qquad
    \ket{oo}\equiv\ket {o}_L\ket {o}_R,
\end{equation}
and then derive the effective model considering the coupling $\Gamma$ in Eq.~\eqref{eq:Htunn} by integrating out the excited states, which gives an effective Hamiltonian spanned by $\left\{ \ket{ee}, \ket{oo} \right\}$ (see Appendix~\ref{app:effective_model} for details)
\begin{equation}\label{eq:effective}
    H =  \frac{E_{+}}{2} \mathds{1}  + \frac{E_{-}}{2}\sigma_z + \frac{\Gamma}{2}\sigma_x,
\end{equation}
where $E_{\pm} \equiv E_{ee}\pm E_{oo}$, with $E_{ee} = E_{e,L}+E_{e,R}$ and $E_{oo} = E_{o,L}+E_{o,R}$ being the total energies of the two chains; in particular, $E_{-}=\varepsilon_L+\varepsilon_R$, where $\varepsilon_a=E_{e,a}-E_{o,a}$ is the splitting Eq.~\eqref{eq:epsilon} of chain $a$.

\section{Results}\label{sec:results}
With the effective Hamiltonian Eq.~\eqref{eq:effective}, we can now study the Rabi and Ramsey protocols of the double $N$-site quantum-dot system. 
As an example, we numerically choose to compute the $N=2$ and $N=3$ results, and compare their coherence times as the number of sites increases. 
To make the comparison on equal footing, we choose the microscopic parameters (i.e., $\mu_i$, $t_i$, and $\Delta_i$) such that the effective $\sigma_x$ rotation strength $\Gamma/2$ and $\sigma_z$ rotation strength $E_{-}/2$ are identical for both $N=2$ and $N=3$ systems in the pristine chains, while keeping the same disorder strength over each microscopic parameter, for both systems.
\subsection{Rabi oscillation protocol}\label{sec:rabi}
\begin{figure}[htbp]
    \centering
    \includegraphics[width=3.4in]{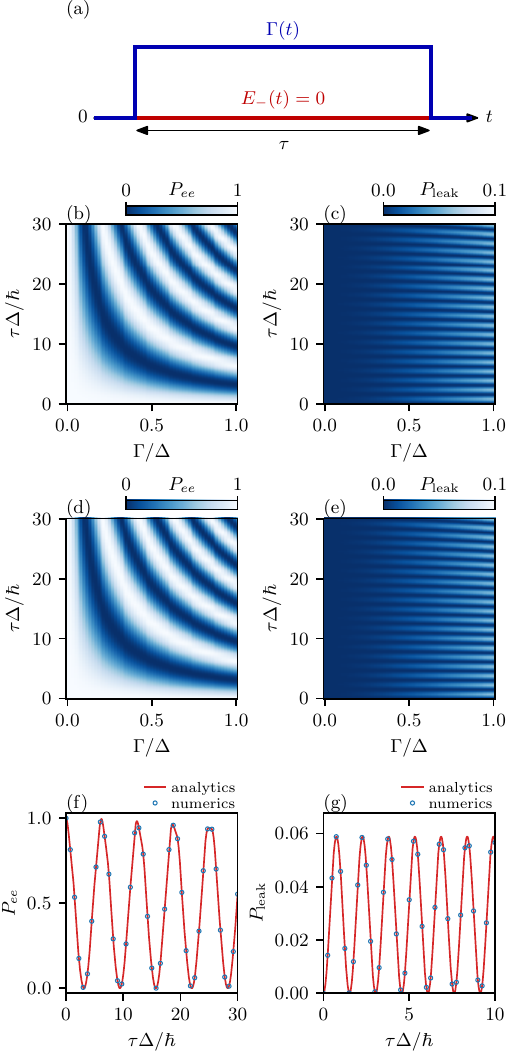}
    \caption{Rabi oscillations in the pristine limit. 
    (a) Pulse profile; 
    (b,c) $P_{ee}$ and $P_{\text{leak}}$ after the pulse versus $\Gamma/\Delta$ and $\tau\Delta/\hbar$ for $N=2$;
    (d,e) the same for $N=3$;
    (f,g) line cuts of (d,e) at $\Gamma/\Delta=1$, with circles being numerical results, and lines being analytical results in Eqs.~\eqref{eq:P_ee_rabi} and \eqref{eq:P_leak_rabi}.}
    \label{fig:pristine_rabi}
\end{figure}

To drive the Rabi oscillation between the two states $\ket{ee}$ and $\ket{oo}$, we initialize the system in the state $\ket{ee}$, and then turn on tunneling $\Gamma$ for a time $\tau$, and finally measure the probability of the system being in the state $\left\{ \ket{ee},\ket{oo} \right\}$, as shown in Fig.~\ref{fig:pristine_rabi}(a). 

For both $N=2$ and $N=3$ systems, since $\Gamma$ is the only control knob for the Rabi oscillation, we keep $\Gamma$ identical for both systems.   
We numerically calculate the probability of finding the system in the state $\ket{ee}$, $P_{ee}$, and the leakage probability $P_{\text{leak}}$ to the excited states, using the full Hamiltonian Eq.~\eqref{eq:Htot} as a function of time, as shown in Fig.~\ref{fig:pristine_rabi}(b,c) for $N=2$ and Fig.~\ref{fig:pristine_rabi}(d,e) for $N=3$.
We find that the Rabi oscillation is almost identical for both systems in the pristine limit as expected, as they share the same $\Gamma$, and $E_-$, which depends on the site numbers as in Eq.~\eqref{eq:epsilon}, is set to zero. So the two systems ($N=2$ and $N=3$) are essentially the same in the pristine limit at the sweet spot. 

In fact, the Rabi oscillation for arbitrary $N\geq2$ can be solved analytically in the pristine limit (see Appendix~\ref{app:rabi_analytical}), with the probability of finding $\ket{ee}$ being
\begin{equation}\label{eq:P_ee_rabi}
    P_{ee}(\tau)=\cos^2\left(\frac{\Gamma}{2}\tau/\hbar\right)\left(1-P_{\text{leak}}(\tau)\right),
\end{equation}
where the leakage probability to the bulk excited states is
\begin{equation}\label{eq:P_leak_rabi}
    \begin{split}
        P_{\text{leak}}(\tau)&=\sin^2\left(\sqrt{\frac{\Gamma^2}{4}+4\Delta^2}\,\tau/\hbar\right)\frac{\Gamma^2}{\Gamma^2+16\Delta^2}\\
        &\approx\frac{\Gamma^2}{16\Delta^2}\sin^2\left(2\Delta\tau/\hbar\right).
    \end{split}
\end{equation}
The line cuts in Fig.~\ref{fig:pristine_rabi}(f,g) show the consistency between the numerical and analytical results in Eqs.~\eqref{eq:P_ee_rabi} and \eqref{eq:P_leak_rabi}.
Since the Rabi oscillation probability Eq.~\eqref{eq:P_ee_rabi} and the leakage probability Eq.~\eqref{eq:P_leak_rabi} do not depend on the number of sites $N$ in the chain, we conclude that the Rabi oscillation is agnostic to the number of sites $N$ in the pristine limit.

\subsection{Ramsey oscillation protocol}\label{sec:ramsey}
\begin{figure}[htbp]
    \centering
    \includegraphics[width=3.4in]{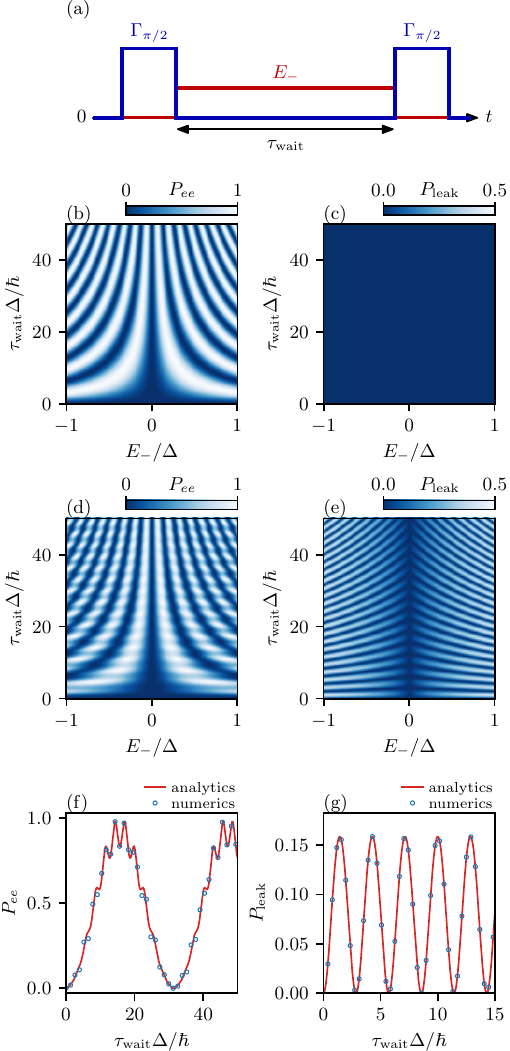}
    \caption{Ramsey oscillations in the pristine limit.
    (a) Pulse sequence: two $\pi/2$ pulses of amplitude $\Gamma_{\pi/2}$ and duration $\tau_{\pi/2}$ enclose the free evolution of duration $\tau_{\text{wait}}$ at the detuning Eqs.~\eqref{eq:ramsey_detuning_N2} and \eqref{eq:ramsey_detuning_N3}-\eqref{eq:calibration} for $N=2$ and $N=3$, respectively; 
    (b,c) $P_{ee}$ and $P_{\text{leak}}$ after the sequence versus $E_-/\Delta$ and $\tau_{\text{wait}}\Delta/\hbar$ for $N=2$;
    (d,e) the same for $N=3$;
    (f,g) line cuts of (d,e) at $E_-/\Delta=0.2$ with circles being numerical results, and lines being analytical results in Eqs.~\eqref{eq:P_ee_ramsey_main} and \eqref{eq:P_leak_ramsey_main}.
    Here, $\Gamma_{\pi/2}=0.2\Delta$;
    }
    \label{fig:pristine_ramsey}
\end{figure}

To drive the Ramsey oscillation, we initialize the system in the state $\ket{ee}$, apply a $\pi/2$ pulse by turning on the tunneling $\Gamma_{\pi/2}$ for a time $\tau_{\pi/2}=\frac{\pi\hbar}{2\Gamma_{\pi/2}}$, which rotates the qubit by $\pi/2$ about the $x$ axis according to Eq.~\eqref{eq:effective} with $E_-=0$, then let the system evolve freely for a time $\tau_{\text{wait}}$ with $\Gamma=0$ and a finite splitting $E_-$, which rotates the qubit about the $z$ axis, and finally apply a second identical $\pi/2$ pulse before measuring the probability of the system being in the state $\left\{ \ket{ee},\ket{oo} \right\}$~\cite{pan2025rabi}, as shown in Fig.~\ref{fig:pristine_ramsey}(a).
Both chains stay at the sweet spot Eq.~\eqref{eq:sweetspot} during the two pulses, so that $E_-=0$ there, and are detuned from it only during the wait.

The $\sigma_z$ rotation thus requires detuning the chains away from the sweet spot, where the protocols for the $N=2$ and $N=3$ systems differ.
For $N=2$, Eq.~\eqref{eq:epsilon_N2} has a contribution linear in $\nu_1$, so a single bond detuning on $t_{L_1}$ produces the splitting~\cite{pan2025rabi},
\begin{equation}\label{eq:ramsey_detuning_N2}
    t_{L_1}:\Delta\to\Delta+u^{(2)},
\end{equation}
which gives $E_-=u^{(2)}$ exactly.
For $N=3$, every term of Eq.~\eqref{eq:epsilon_N3} is a product of at least two local detunings, thus any single detuning of $t_i$, $\Delta_i$, or $\mu_i$ creates no splitting. Therefore, we need to tune at least two parameters simultaneously. 
Without loss of generality, we choose to detune the first bond $t_{a_1}$ and the last site chemical potential $\mu_{a_3}$ of each chain $a\in\{L,R\}$ simultaneously, 
\begin{equation}\label{eq:ramsey_detuning_N3}
    t_{a_1}:\Delta\to\Delta+u^{(3)},\; \mu_{a_3}:0\to \text{sgn}(E_-)\,u^{(3)}, \quad a\in\{L,R\},
\end{equation}
such that the first term of Eq.~\eqref{eq:epsilon_N3} is finite.
Here, in order to maintain the same splitting $E_-$ as in $N=2$, we choose the detuning $u^{(3)}$ as (see Appendix~\ref{app:ramsey_N3})
\begin{equation}\label{eq:calibration}
    u^{(3)}=\sqrt{\abs{E_-}\left(\Delta+\frac{\abs{E_-}}{4}\right)}\approx\sqrt{\Delta \abs{E_-}},
\end{equation}
where the last expression holds for $\abs{E_-}\ll\Delta$.

We numerically calculate the probability of finding the system in the state $\ket{ee}$, $P_{ee}$, and the leakage probability $P_{\text{leak}}$ to the excited states after the full sequence, using the full Hamiltonian Eq.~\eqref{eq:Htot} with the detunings $u^{(2)}$ and $u^{(3)}$ calibrated to the same $E_-$ for both systems, as a function of the wait time $\tau_{\text{wait}}$ and the splitting $E_-$, as shown in Fig.~\ref{fig:pristine_ramsey}(b,c) for $N=2$ and Fig.~\ref{fig:pristine_ramsey}(d,e) for $N=3$.
The fringes of $P_{ee}$ in Fig.~\ref{fig:pristine_ramsey}(b,d) are nearly identical, while the leakage differs qualitatively as in Fig.~\ref{fig:pristine_ramsey}(c,e).

For $N=2$ at $\mu_{i}=0$, the detuned bond $\frac{i}{2}\nu_{1}\gamma_{1}\bar{\gamma}_{2}$ of Eq.~\eqref{eq:singlechain} equals $\frac{\nu_{1}}{2}P\,i\bar{\gamma}_{1}\gamma_{2}$ with the parity $P$ of Eq.~\eqref{eq:parity}, so $\ket{e}$ and $\ket{o}$ of Eq.~\eqref{eq:two_site_states} remain exact eigenstates during the wait for any $t_{L_1}$; the bulk is reached only through $\Gamma$ during the two $\pi/2$ pulses, and $P_{\text{leak}}$ in Fig.~\ref{fig:pristine_ramsey}(c) stays at the $10^{-6}$ level of Eq.~\eqref{eq:P_leak_rabi} at $\tau=\tau_{\pi/2}$.

For $N=3$, the detunings of Eq.~\eqref{eq:ramsey_detuning_N3} required for the $\sigma_z$ rotation also couple the qubit to the bulk, so that the wait is no longer leakage free.
The two $\pi/2$ pulses leak only at second order in $\Gamma_{\pi/2}/\Delta$, and we neglect them relative to the leakage during the wait, which is first order in $\abs{E_-}/\Delta$.
During the wait, $\ket{ee}$ remains an exact eigenstate of the evolution while $\ket{oo}$ leaks into the bulk for $E_->0$, and vice versa for $E_-<0$.
After some algebra, we derive the leakage probability (see Appendix~\ref{app:ramsey_analytical} for the derivation)
\begin{equation}\label{eq:P_leak_ramsey_main}
    \begin{split}
        P_{\text{leak}}(\tau_{\text{wait}})&=\frac{1-\left[1-P_{01}(\tau_{\text{wait}})\right]^{2}}{2}=P_{01}-\frac{P_{01}^{2}}{2}\\
        &\approx\frac{\abs{E_-}}{\Delta}\sin^{2}\left(\frac{\left(2\Delta+\abs{E_-}\right)\tau_{\text{wait}}}{2\hbar}\right),
    \end{split}
\end{equation}
where $P_{01}$ is the leakage probability of a single chain [Eq.~\eqref{eq:P01}], and the return probability
\begin{equation}\label{eq:P_ee_ramsey_main}
    P_{ee}(\tau_{\text{wait}})\approx\left[1-P_{01}(\tau_{\text{wait}})\right]\sin^2\left(\frac{E_-\tau_{\text{wait}}}{2\hbar}\right)+\frac{P_{01}^2}{4}.
\end{equation}
The factor $1-P_{01}$, coming from the leakage, contributes to the ripple on the fringes in Fig.~\ref{fig:pristine_ramsey}(f), which runs at a much higher bulk excitation frequency $(2\Delta+\abs{E_-})/\hbar$ of Eq.~\eqref{eq:P01}. 
This fine structure distinguishes Fig.~\ref{fig:pristine_ramsey}(d) from (b). 
The numerical consistency with the analytical results in Eqs.~\eqref{eq:P_ee_ramsey_main} and \eqref{eq:P_leak_ramsey_main} is shown in Fig.~\ref{fig:pristine_ramsey}(f,g).

The significantly higher leakage in $N=3$ compared to $N=2$ is the trade-off for the three-site protection: the absence of first-order terms in Eq.~\eqref{eq:epsilon_N3} suppresses the deliberate splitting as much as the noise-induced one, so the detuning $u^{(3)}\approx\sqrt{\Delta\abs{E_-}}\gg\abs{E_-}=\abs{u^{(2)}}$ required for a given $E_-$ excites the bulk.

\subsection{Coherence time $T_{2,\text{Rabi}}$ and $T_{2,\text{Ramsey}}$}
To study the coherence time of the Rabi oscillation $T_{2,\text{Rabi}}$ and Ramsey oscillation $T_{2,\text{Ramsey}}$, we introduce quenched disorder to the microscopic parameters of the two chains.
We consider three types of typical parameters and then numerically solve the full Hamiltonian in Eq.~\eqref{eq:Htot} to calculate each Rabi oscillation and Ramsey oscillation for both $N=2$ and $N=3$ systems to obtain the typical coherence time in Sec.~\ref{sec:Rabi_disorder} and Sec.~\ref{sec:Ramsey_disorder}, respectively. 
The three typical parameters are 
(1) small superconducting gap;
(2) large superconducting gap, both from Table~I of Ref.~\cite{pan2025rabi}; 
and (3) recent experimental parameters from Table~1 of Ref.~\cite{zatelli2026majorana}. 
The disorder in each microscopic parameter $x$ is modeled as a Gaussian distribution $\mathcal{N}(\bar{x},\sigma_{x}^{2})$ with mean $\bar{x}$ and standard deviation $\sigma_{x}$, as documented in Table~\ref{tab:parameters}.
For the experimental parameters, $\Gamma$ is the effective junction coupling mediated by the additional middle normal dot of Ref.~\cite{zatelli2026majorana}, which we derive in Appendix~\ref{app:Gamma_exp}.
The resulting typical coherence times of both protocols are summarized in Table~\ref{tab:comparison}.
Finally, we consider the sweep of microscopic parameters to systematically study the change of the coherence time from $N=2$ to $N=3$ in Sec.~\ref{sec:sweep}.

\begin{table}[htbp]
    \centering
    \small
    \caption{Simulation parameters in $\mu$eV}
    \label{tab:parameters}
    \begin{tabular}{lccc}
        \toprule
        Parameter & Small gap~\cite{pan2025rabi} & Large gap~\cite{pan2025rabi} & Experiment~\cite{zatelli2026majorana} \\
        \midrule
        $\mu_{a_i}$  & $\mathcal{N}(0,3^2)$  & $\mathcal{N}(0,0.3^2)$  & $\mathcal{N}(0,1.5^2)$  \\
        $t_{a_i}$ & $\mathcal{N}(12,0.0158^2)$  & $\mathcal{N}(38,0.05^2)$  & $\mathcal{N}(20,0.07^2)$  \\
        $\Delta_{a_i}$  & $\mathcal{N}(12,0)$  & $\mathcal{N}(38,0)$  & $\mathcal{N}(20,0.07^2)$  \\
        $\Gamma$  & $\mathcal{N}(5,0.05^2)$  & $\mathcal{N}(5,0.05^2)$  & $\mathcal{N}(0.433,0.017^2)$  \\
        $\Gamma_{\pi/2}$  & 0.5  & 0.5  & 0.5  \\
        $E_-$ ($=u^{(2)}$) & 2 & 2 & 2 \\
        $u^{(3)}$ & 5.00 & 8.77 & 6.40 \\
        \bottomrule
    \end{tabular}
\end{table}

\begin{table*}[htbp]
    \centering
    \small
    \caption{Rabi and Ramsey oscillations for $N=2$ and $N=3$ at the parameters of Table~\ref{tab:parameters}. The measured column is from Ref.~\cite{zatelli2026majorana}; all other columns are simulated in Figs.~\ref{fig:disorder_rabi} and~\ref{fig:disorder_ramsey}, with $f$, $T_2$, $2A$, and $\beta$ from the fit Eq.~\eqref{eq:fit}, $Q$ from Eq.~\eqref{eq:quality}, and $P_{\text{leak}}$ from Eq.~\eqref{eq:leak_average}.}
    \label{tab:comparison}
    \begin{tabular}{llccccccc}
\toprule
 & & \multicolumn{2}{c}{Small gap} & \multicolumn{2}{c}{Large gap} & \multicolumn{3}{c}{Experiment} \\
\cmidrule(lr){3-4}\cmidrule(lr){5-6}\cmidrule(lr){7-9}
Protocol & Quantity & $N=2$ & $N=3$ & $N=2$ & $N=3$ & $N=2$ (meas.) & $N=2$ & $N=3$ \\
\midrule
Rabi & $f$ (GHz) & 1.1966 & 1.1980 & 1.2090 & 1.2091 & 0.1067(4) & 0.1071 & 0.1045 \\
 & $T_2$ (ns) & 10.42 & 10.78 & 18.91 & 17.53 & 26(2) & 37.02 & 53.75 \\
 & $Q$ & 78.3 & 81.2 & 143.6 & 133.2 & 17(1) & 24.9 & 35.3 \\
 & $2A$ & 0.962 & 0.973 & 0.999 & 1.008 & -- & 0.931 & 1.006 \\
 & $\beta$ & 1.66 & 1.66 & 2.02 & 1.88 & 1 (fixed) & 1.29 & 1.86 \\
 & $P_{\text{leak}}$ & $3.7\times10^{-2}$ & $4.1\times10^{-2}$ & $5.7\times10^{-4}$ & $5.8\times10^{-4}$ & $\approx0.05$\footnote{Dominated by readout errors according to Ref.~\cite{zatelli2026majorana}, hence not comparable to the simulated bulk leakage.} & $2.8\times10^{-3}$ & $3.1\times10^{-3}$ \\
\midrule
Ramsey & $f$ (GHz) & 0.4836 & 0.4836 & 0.4836 & 0.4850 & -- & 0.4836 & 0.4836 \\
 & $T_2$ (ns) & 1.77 & 1.16 & 10.12 & 20.28 & -- & 3.55 & 2.94 \\
 & $Q$ & 5.4 & 3.5 & 30.8 & 61.8 & -- & 10.8 & 8.9 \\
 & $2A$ & 0.328 & 0.885 & 0.921 & 0.988 & -- & 0.788 & 0.952 \\
 & $\beta$ & 0.92 & 1.80 & 1.97 & 1.76 & -- & 1.49 & 1.97 \\
 & $P_{\text{leak}}$ & $2.9\times10^{-2}$ & $1.0\times10^{-1}$ & $3.2\times10^{-5}$ & $2.5\times10^{-2}$ & -- & $2.8\times10^{-3}$ & $4.8\times10^{-2}$ \\
\bottomrule
\end{tabular}

\end{table*}

\subsubsection{Rabi oscillation with disorder}\label{sec:Rabi_disorder}

\begin{figure*}[ht]
    \centering
    \includegraphics[width=7in]{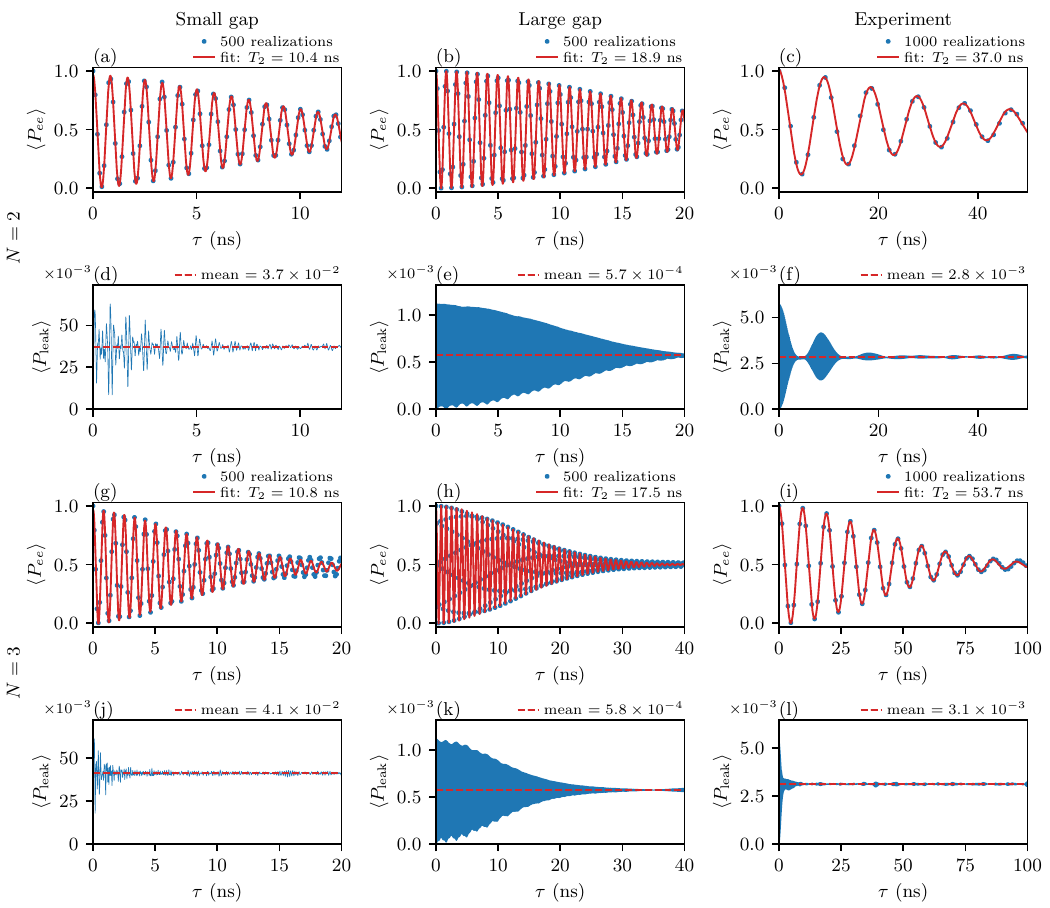}
    \caption{Rabi oscillation with disorder. The first two rows are for $P_{ee}$ and $P_{\text{leak}}$ for $N=2$, and the last two rows are the corresponding values for $N=3$. The first column is for small gap parameters, the second column is for large gap parameters, and the third column is for experimental parameters. The parameters are listed in Table~\ref{tab:parameters}. }
    \label{fig:disorder_rabi}
\end{figure*}

We present the Rabi oscillation with disorder for both the double two-site and double three-site quantum-dot chains in Fig.~\ref{fig:disorder_rabi}. 

We take the ensemble average $\expval{P_{ee}}$ over 500 realizations (for small and large gap parameters) and 1000 realizations (for experimental parameters) of the disorder, and fit it with the following function~\cite{pan2025rabi}
\begin{equation}\label{eq:fit}
    \expval{P_{ee}(\tau)}=P_{0}+A\cos\left(2\pi f\tau+\phi_{0}\right)\exp\left[-\left(\tau/T_{2}\right)^{\beta}\right],
\end{equation}
where $2A$ is the visibility and $\beta$ is the damping exponent. 
From the fitted $f$ and $T_2$, we obtain the quality factor
\begin{equation}\label{eq:quality}
    Q = 2\pi f T_{2},
\end{equation}
which counts the coherent phase accumulated before the oscillation envelope decays to $1/e$, and the time-averaged leakage
\begin{equation}\label{eq:leak_average}
    \expval{P_{\text{leak}}}=\frac{1}{\tau_{\max}}\int_{0}^{\tau_{\max}}\left[1-\expval{P_{ee}(\tau)}-\expval{P_{oo}(\tau)}\right]\dd\tau,
\end{equation}
where $P_{oo}$ is the probability of $\ket{oo}$ and $\tau_{\max}$ is the length of the simulated window.
In Figs.~\ref{fig:disorder_rabi} and~\ref{fig:disorder_ramsey}, for better visibility, the markers show $\expval{P_{ee}}$ subsampled to eight points per oscillation period, i.e., one to two orders of magnitude coarser than the simulation time step; nevertheless, Eq.~\eqref{eq:fit} is fitted to the full data.

We find that the fitted $T_{2,\text{Rabi}}$ of Eq.~\eqref{eq:fit} for both $N=2$ and $N=3$ are qualitatively similar in the small and large gap parameters, while the experimental parameters show an improvement of $T_{2,\text{Rabi}}$ by a factor of 1.4 for $N=3$ over $N=2$. 
The leakage $\expval{P_{\text{leak}}}$ of Eq.~\eqref{eq:leak_average} remains almost the same magnitude for both systems.

\subsubsection{Ramsey oscillation with disorder}\label{sec:Ramsey_disorder}
\begin{figure*}[ht]
    \centering
    \includegraphics[width=7in]{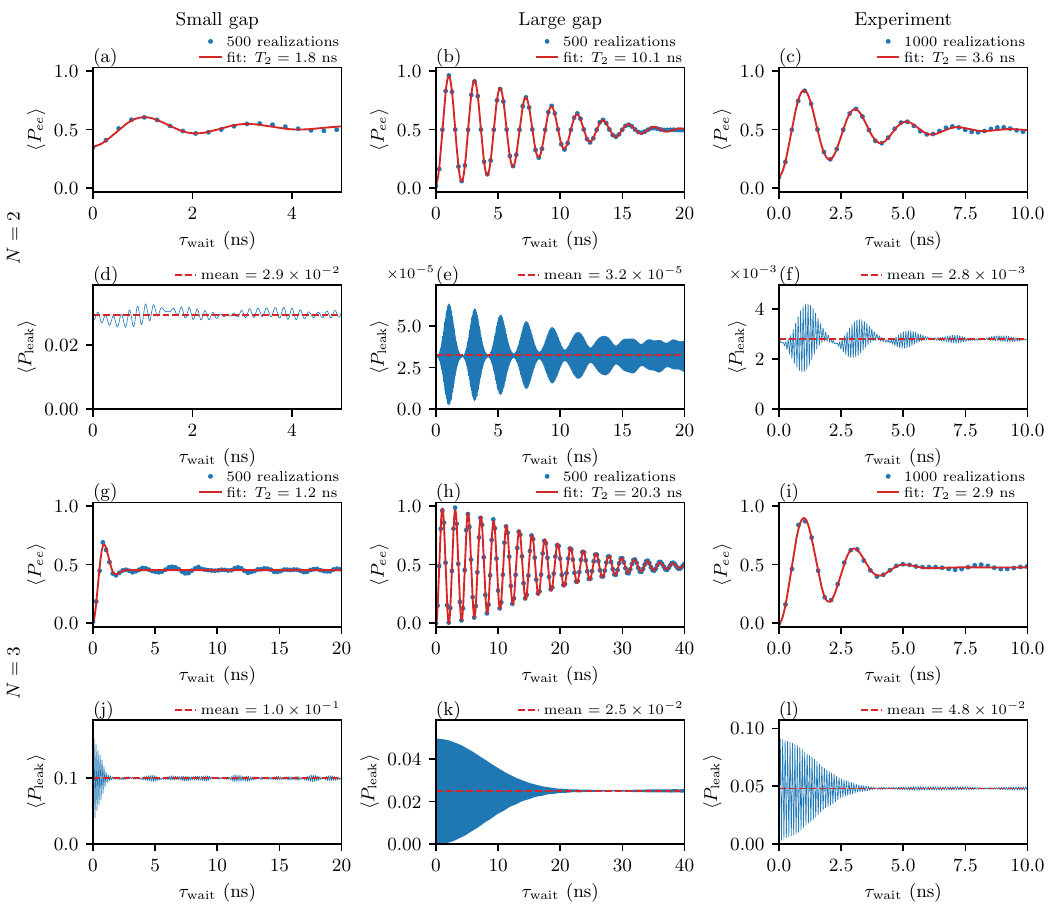}
    \caption{Ramsey oscillation with disorder. The first two rows are for $P_{ee}$ and $P_{\text{leak}}$ for $N=2$, and the last two rows are the corresponding values for $N=3$. The first column is for small gap parameters, the second column is for large gap parameters, and the third column is for experimental parameters. The parameters are listed in Table~\ref{tab:parameters}. }
    \label{fig:disorder_ramsey}
\end{figure*}

In Fig.~\ref{fig:disorder_ramsey}, we present the Ramsey oscillation with disorder for both the double two-site and double three-site quantum-dot chains.
We find that the large gap parameters show an improvement of the fitted $T_{2, \text{Ramsey}}$ of Eq.~\eqref{eq:fit} by a factor of 2; however, the small gap parameters and the experimental parameters show quantitatively similar $T_{2, \text{Ramsey}}$.
The leakage $\expval{P_{\text{leak}}}$ of Eq.~\eqref{eq:leak_average}, on the other hand, shows a significant degradation from $N=2$ to $N=3$ for all three types of parameters, which is consistent with the pristine limit results in Fig.~\ref{fig:pristine_ramsey}. 

\subsubsection{Sweep of microscopic parameters}\label{sec:sweep}

Finally, from the above three typical sets of parameters, we find that the coherence time does not always show an improvement when the number of sites increases from $N=2$ to $N=3$.
Therefore, to systematically study it and avoid bias in the choice of parameters, we sweep the microscopic parameters in Table~\ref{tab:parameters} to see the ratio of $T_2$ from $N=2$ to $N=3$.

\begin{figure*}[htbp]
    \centering
    \includegraphics[width=7in]{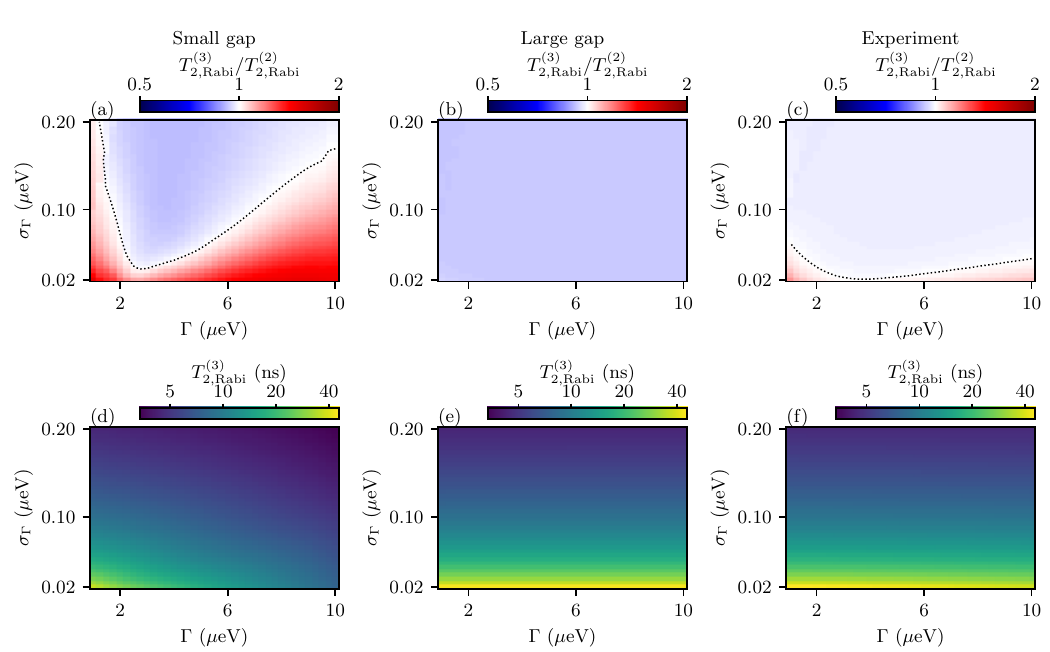}
    \caption{
    Rabi coherence times fitted by Eq.~\eqref{eq:fit}, for the small gap (left column), large gap (middle column), and experimental (right column) parameters of Table~\ref{tab:parameters}.
    (a-c) $T_{2,\text{Rabi}}^{(3)}/T_{2,\text{Rabi}}^{(2)}$ in the log scale versus the tunneling $\Gamma$ and its disorder $\sigma_\Gamma$; white color and the dotted contours mark the equality of two coherence times. 
    (d-f) $T_{2,\text{Rabi}}^{(3)}$ in the log scale over the same sweep.
    Each point uses $500$ (small and large gap) or $1000$ (experimental) disorder realizations. All other parameters are as in Table~\ref{tab:parameters}.}
    \label{fig:T2_ratio_rabi}
\end{figure*}

Figure~\ref{fig:T2_ratio_rabi}(a-c) shows the ratio $T_{2,\text{Rabi}}^{(3)}/T_{2,\text{Rabi}}^{(2)}$ of the fitted Rabi coherence times of $N=3$ and $N=2$ as a function of the tunneling strength $\Gamma$ and the junction disorder $\sigma_\Gamma$ for the small gap parameters (first column in Table~\ref{tab:parameters}), large gap parameters (second column in Table~\ref{tab:parameters}), and experimental parameters (third column in Table~\ref{tab:parameters}).
We find that it does not show a universal improvement: for the small gap parameters, it shows a U-shape: around $\Gamma=3\ueV$, the improvement is minimal, whereas for larger or smaller $\Gamma$, we see an improvement of $T_{2,\text{Rabi}}$ up to a factor of $1.5$.
However, for the large gap parameters, the improvement is minimal at the order of $\mathcal{O}(1)$.
The experimental parameters show a similar trend to the small gap parameters, with a U-shaped contour.
The corresponding $T_{2,\text{Rabi}}^{(3)}$ is shown in Fig.~\ref{fig:T2_ratio_rabi}(d-f) for the same range of $\Gamma$ and $\sigma_\Gamma$.
The different trends of the ratio are due to the competition of the three different dephasing channels as discussed in Sec.~\ref{sec:discussion}.

\begin{figure*}[htbp]
    \centering
    \includegraphics[width=7in]{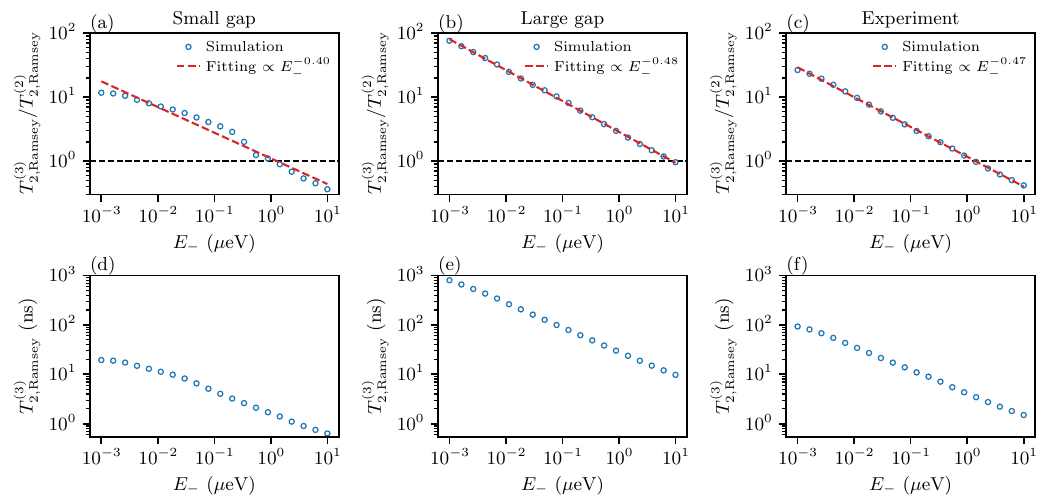}
    \caption{
    Ramsey coherence times fitted by Eq.~\eqref{eq:fit}, for the small gap (left column), large gap (middle column), and experimental (right column) parameters of Table~\ref{tab:parameters}.
    (a-c) $T_{2,\text{Ramsey}}^{(3)}/T_{2,\text{Ramsey}}^{(2)}$ in the log-log scale versus the splitting $E_-$, with $u^{(3)}$ calibrated to $E_-$ by Eq.~\eqref{eq:calibration}; the horizontal black dashed line marks ratio being unity, and the red dashed line is the power-law fit;
    (d-f) $T_{2,\text{Ramsey}}^{(3)}$ in the log-log scale over the same sweep.
    Each point uses $500$ (small and large gap) or $1000$ (experimental) disorder realizations. All other parameters are as in Table~\ref{tab:parameters}.}
    \label{fig:T2_ratio_ramsey}
\end{figure*}

Similarly, Fig.~\ref{fig:T2_ratio_ramsey}(a-c) shows the ratio of the Ramsey coherence times $T_{2,\text{Ramsey}}^{(3)}/T_{2,\text{Ramsey}}^{(2)}$ as a function of the splitting $E_-$ for small, large gap parameters, and the experimental parameters, with all other parameters as in Table~\ref{tab:parameters}.
The Ramsey coherence time improves from $N=2$ to $N=3$ only below a crossover splitting $E_-=1.2$, $9.4$, and $1.4\ueV$ for the three sets; below it, the ratio grows as a power law $E_-^{-p}$ with fitted exponents $p=0.40$, $0.48$, and $0.47$, close to the analytical value $p=-1/2$ as shown in  Sec.~\ref{sec:ramsey_dephasing}, and reaches $12$, $76$, and $27$ at $E_-=10^{-3}\ueV$.
The corresponding $T_{2,\text{Ramsey}}^{(3)}$ is shown in Fig.~\ref{fig:T2_ratio_ramsey}(d-f) for the same range of $E_-$.

\section{Discussion}\label{sec:discussion}
In Figs.~\ref{fig:disorder_rabi}-\ref{fig:T2_ratio_ramsey}, we find that the $T_{2,\text{Rabi}}$ and $T_{2,\text{Ramsey}}$ do not show a universal improvement from $N=2$ to $N=3$. Here we provide a heuristic analytical study focusing on the dephasing channels for this observation in both Rabi oscillation in Sec.~\ref{sec:rabi_dephasing} and Ramsey oscillation in Sec.~\ref{sec:ramsey_dephasing}, respectively.

\subsection{Three competing dephasing channels in Rabi oscillation}\label{sec:rabi_dephasing}
For the Rabi oscillation, there are three competing dephasing channels playing a role here, and the eventual coherence time is determined by which channel dominates the dephasing. 
The three channels are
(1) $T_{2, \text{Rabi},\Gamma}^{(N)}$ controlled by the disorder in tunneling $\Gamma$ between the two chains, which is not sensitive to the bulk properties of the chains, and thus the increased number of sites barely improves the coherence time; 
(2) $T_{2, \text{Rabi},E_-}^{(N)}$ controlled by the disorder in $t_i$, $\Delta_i$, and $\mu_i$ which leads to the detuning from the sweet spot, which is suppressed by the three-site protection, and 
(3) $T_{2, \text{Rabi},\Gamma_{\text{eff}}}^{(N)}$ controlled by the disorder in $t_i$, $\Delta_i$, and $\mu_i$ which renormalizes the interchain tunneling which is also suppressed by the three-site protection.
Here, $N$ denotes the number of sites in each chain.
We discuss each of the three dephasing channels in the following Sec.~\ref{sec:interchain_tunneling}, Sec.~\ref{sec:interchain_splitting}, and Sec.~\ref{sec:renormalization_tunneling}, respectively. 

\subsubsection{Dephasing channel 1: intrinsic interchain tunneling}\label{sec:interchain_tunneling}
The first dephasing channel is purely induced by $\sigma_\Gamma$, the intrinsic fluctuation of the interchain tunneling $\Gamma$.
The single-trajectory return probability $P_{ee}$ is given by Eq.~\eqref{eq:P_ee_rabi}. 
Averaging its oscillatory part over the Gaussian noise of $\Gamma\sim\mathcal{N}(\bar\Gamma,\sigma_\Gamma^2)$, $\expval{\cos\left(\frac{\Gamma t}{\hbar}\right)}=\cos\left(\frac{\bar\Gamma t}{\hbar}\right)\exp\left[-\frac{\sigma_\Gamma^2t^2}{2\hbar^2}\right]$, and matching this damping envelope to $\exp[-(t/T_2)^\beta]$ in Eq.~\eqref{eq:fit}, we have $\beta=2$ and  
\begin{equation}\label{eq:rabi_theory}
    T_{2,\text{Rabi},\Gamma}^{(N)}=\frac{\sqrt2\hbar}{\sigma_\Gamma}.
\end{equation} 
for both $N=2$ and $N=3$.
Therefore, the intrinsic interchain tunneling dephasing channel is agnostic to the system lengths.

\subsubsection{Dephasing channel 2: intrachain splitting}\label{sec:interchain_splitting}
The second dephasing channel is induced by the disorder in $t_{a_i}$, $\Delta_{a_i}$, and $\mu_{a_i}$, detuning both chains from the sweet spot to create a random splitting.
With the approximation of zero leakage and a deterministic $\Gamma$, this energy splitting leads to a $\sigma_z$ rotation in Eq.~\eqref{eq:effective}.

Therefore, the return probability becomes
\begin{equation}\label{eq:P_ee_rabi_detuned}
    \begin{split}
        P_{ee}(t)&=1-\frac{\Gamma^2}{\Gamma^2+E_-^2}\sin^2\left(\frac{\sqrt{\Gamma^2+E_-^2}\,t}{2\hbar}\right)\\
        &\approx\cos^2\left(\frac{\Gamma t}{2\hbar}+\frac{E_-^2t}{4\Gamma\hbar}\right),
    \end{split}
\end{equation}
in the limit of $\abs{E_-}\ll\Gamma$.~\footnote{The prefactor only reduces the oscillation amplitude by the static factor $1-E_-^2/\Gamma^2$, whereas the Rabi frequency shifts to $\sqrt{\Gamma^2+E_-^2}\approx\Gamma+E_-^2/2\Gamma$, serving as the second-order phase drift $E_-^2t/2\Gamma\hbar$, unbounded in $t$, that will dephase the oscillation.}
Approximating the non-Gaussian $E_-$ by a zero-mean Gaussian variable $E_- \sim \mathcal{N}(0,\sigma_{E_-}^2)$, the disorder average is $\expval{P_{ee}(t)} = \frac{1}{2} + \frac{1}{2} \abs{\expval{ e^{iE_-^2 t/2\Gamma\hbar}}} \cos\left(\frac{\Gamma t}{\hbar}+\varphi(t)\right)$, where the phase $\varphi(t) = \arg \expval{ e^{iE_-^2 t/2\Gamma\hbar}}$. 
This disorder-averaged oscillatory part gives the algebraic envelope $\abs{\expval{e^{iE_-^2t/2\Gamma\hbar}}}=\left[1+\left({\sigma_{E_-}^2t}/{\Gamma\hbar}\right)^2\right]^{-1/4}$, which is not of the standard exponential form of Eq.~\eqref{eq:fit}; nevertheless, we identify an operational $T_2$ with its $1/e$ time as
\begin{equation}\label{eq:rabi_eps}
    T_{2,\text{Rabi},E_-}^{(N)}=\sqrt{e^4-1}\,\frac{\Gamma\hbar}{\sigma_{E_-}^2}.
\end{equation}

The dependence on $N$ enters through $\sigma_{E_-}^2$.
At the sweet spot, $\nu_{a_i}$ and $\mu_{a_i}$ are independent zero-mean Gaussians with variances $\sigma_\nu^2=\sigma_t^2+\sigma_\Delta^2$ and $\sigma_\mu^2$, and $s_{a_i}\approx2\Delta$ to leading order.
Since the two chains are independent and the terms of Eqs.~\eqref{eq:epsilon_N2} and \eqref{eq:epsilon_N3} are mutually uncorrelated,
\begin{equation}\label{eq:E_minus_variance_N2}
        \sigma_{E_-}^2(N=2)=2\left[\sigma_\nu^2+\frac{\sigma_\mu^4}{(2\Delta)^2}\right],
\end{equation}
and
\begin{equation}\label{eq:E_minus_variance_N3}
        \sigma_{E_-}^2(N=3)=2\left[\frac{2\sigma_\nu^2\sigma_\mu^2}{(2\Delta)^2}+\frac{\sigma_\mu^6}{(2\Delta)^4}\right],
\end{equation}
where the factor of $2$ counts the two chains.
Since $\Gamma$ is identical for both $N$, Eqs.~\eqref{eq:rabi_eps}, \eqref{eq:E_minus_variance_N2}, and \eqref{eq:E_minus_variance_N3} give
\begin{equation}\label{eq:rabi_eps_ratio}
    \frac{T_{2,\text{Rabi},E_-}^{(3)}}{T_{2,\text{Rabi},E_-}^{(2)}}=\frac{(2\Delta)^2}{\sigma_\mu^2}\,\frac{(2\Delta)^2\sigma_\nu^2+\sigma_\mu^4}{2(2\Delta)^2\sigma_\nu^2+\sigma_\mu^4},
\end{equation}
which lies between $2\Delta^2/\sigma_\mu^2$ and $4\Delta^2/\sigma_\mu^2$, reflecting the additional weak bond $\mu_{a_i}/s_{a_j}\sim\sigma_\mu/2\Delta$ that the third site adds to every path of Eq.~\eqref{eq:epsilon}.
Therefore, the intrachain splitting dephasing channel is suppressed by the third site by a factor of order $(\Delta/\sigma_\mu)^2\gg1$.

\subsubsection{Dephasing channel 3: renormalization of interchain tunneling}\label{sec:renormalization_tunneling}
The third dephasing channel is also induced by the disorder in $t_{a_i}$, $\Delta_{a_i}$, and $\mu_{a_i}$, detuning from the sweet spot to create the delocalized end Majorana modes, which can further renormalize the interchain tunneling $\Gamma$. 
Unlike the second dephasing channel which introduces an unwanted $\sigma_z$ rotation in the Rabi oscillation, this third channel introduces an additional $\sigma_x$ rotation on top of the external $\Gamma$.

At the sweet spot, the end modes adjacent to the junction are exactly localized at $\bar{\gamma}_{L_N}$ and $\gamma_{R_1}$;
however, when away from the sweet spot, both end Majoranas penetrate into the bulk as~\footnote{
To first order in the weak bonds, the end modes become $\tilde{\bm{\gamma}}_E=\bm{\gamma}_E-\left(K_S^{-1}K_{BE}\right)^\intercal\bm{\gamma}_B$ (see Eq.~\eqref{eq:Kblocks}), i.e., each tail traverses one weak bond and one strong bond (see Fig.~\ref{fig:paths}(a)).}
\begin{equation}\label{eq:mode_tails}
    \begin{split}
        \tilde{\gamma}_{1}&\propto\gamma_{1}-\frac{\mu_{1}}{s_{1}}\gamma_{2}+\frac{\nu_{1}}{s_{2}}\gamma_{3},\\
        \tilde{\bar{\gamma}}_{3}&\propto\bar{\gamma}_{3}-\frac{\mu_{3}}{s_{2}}\bar{\gamma}_{2}+\frac{\nu_{2}}{s_{1}}\bar{\gamma}_{1},
    \end{split}
\end{equation}
with the normalization factor $\left(1+\mu_{3}^2/s_{2}^2+\nu_{2}^2/s_{1}^2\right)^{-1/2}\approx 1-\frac{1}{2}\left(\mu_{3}^2/s_{2}^2+\nu_{2}^2/s_{1}^2\right)$ for $\tilde{\bar{\gamma}}_{3}$ and similarly for $\tilde{\gamma}_{1}$ for $N=3$.
The delocalized Majorana modes renormalize the effective tunneling as $\Gamma_{\text{eff}}\equiv2\abs{\mel{oo}{\mathcal{H}_{\text{tunn}}}{ee}}$, leading to
\begin{equation}\label{eq:gamma_eff_N3}
        \Gamma_{\text{eff}}^{(3)}\approx\Gamma\left(1-\frac{\mu_{L_3}^2+\nu_{L_2}^2+\mu_{R_1}^2+\nu_{R_1}^2}{2(2\Delta)^2}-\frac{\nu_{L_1}\nu_{R_2}}{(2\Delta)^2}\right).
\end{equation}

For $N=2$, the end modes are $\tilde{\gamma}_{1}\propto\gamma_{1}-\frac{\mu_{1}}{s_{1}}\gamma_{2}$ and $\tilde{\bar{\gamma}}_{2}\propto\bar{\gamma}_{2}-\frac{\mu_{2}}{s_{1}}\bar{\gamma}_{1}$, whereas $\nu_{1}$ couples $\gamma_{1}$ and $\bar{\gamma}_{2}$ directly and enters only $E_-$ [Eq.~\eqref{eq:epsilon_N2}], leading to a renormalized tunneling
\begin{equation}\label{eq:gamma_eff_N2}
    \Gamma_{\text{eff}}^{(2)}\approx\Gamma\left( 1-\frac{\mu_{L_2}^2+\mu_{R_1}^2+2\mu_{L_1}\mu_{R_2}}{2(2\Delta)^2}\right) .
\end{equation}

Averaged over the quenched disorder, $\delta\Gamma_{\text{eff}}^{(N)}\equiv \Gamma_{\text{eff}}^{(N)}-\Gamma$ is quadratic in the Gaussian variables $\mu_{a_i}$ and $\nu_{a_i}$, so the damping envelope of $\expval{\cos\left(\Gamma_{\text{eff}}t/\hbar\right)}$ is again algebraic, and we again identify $T_2$ with its $1/e$ time (see Appendix~\ref{app:geff_average}).
Given $\sigma_\nu\ll\sigma_\mu$, true for all three parameter sets in Table~\ref{tab:parameters}, the $\nu_{a_i}$ terms of Eq.~\eqref{eq:gamma_eff_N3} are negligible, and the disorder averages of Eqs.~\eqref{eq:gamma_eff_N3} and \eqref{eq:gamma_eff_N2} give
\begin{equation}\label{eq:rabi_perp}
    \begin{split}
        \abs{\expval{e^{i\delta\Gamma_{\text{eff}}^{(3)}t/\hbar}}}&=\left(1+x^2\right)^{-1/2},\\
        \abs{\expval{e^{i\delta\Gamma_{\text{eff}}^{(2)}t/\hbar}}}&=\left(1+x^2\right)^{-1},
    \end{split}
\end{equation}
with $x\equiv\Gamma\sigma_\mu^2t/[(2\Delta)^2\hbar]$, whose $1/e$ times are
\begin{equation}\label{eq:T2_geff}
    \begin{split}
        T_{2,\text{Rabi},\Gamma_{\text{eff}}}^{(3)}&=\sqrt{e^2-1}\,\frac{(2\Delta)^2\hbar}{\Gamma\sigma_\mu^2},\\
        T_{2,\text{Rabi},\Gamma_{\text{eff}}}^{(2)}&=\sqrt{e-1}\,\frac{(2\Delta)^2\hbar}{\Gamma\sigma_\mu^2}.
    \end{split}
\end{equation}

Therefore, the third site improves this channel by the fixed factor 
\begin{equation}\label{eq:T2_geff_ratio}
    T_{2,\text{Rabi},\Gamma_{\text{eff}}}^{(3)}/T_{2,\text{Rabi},\Gamma_{\text{eff}}}^{(2)}=\sqrt{e+1}\approx1.93.
\end{equation}

\subsubsection{Crossover between the three dephasing channels}\label{sec:rabi_crossover}

\begin{figure*}[ht]
    \centering
    \includegraphics[width=7in]{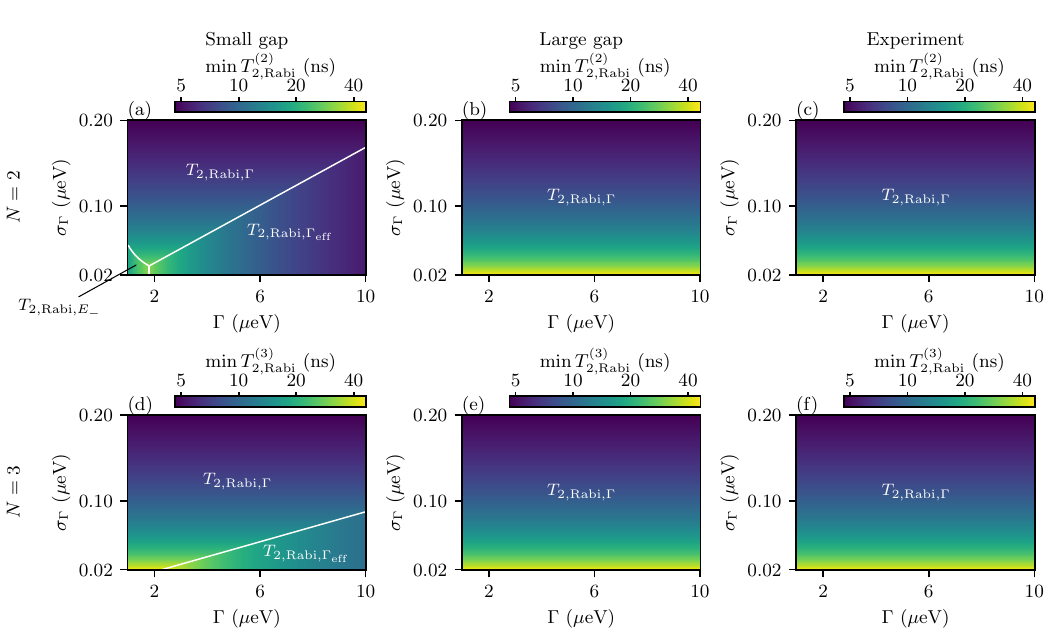}
    \caption{Dephasing channels in the Rabi oscillation. The heuristic coherence time of Eq.~\eqref{eq:rabi_T2_min}, the shortest of the three channel times of Eqs.~\eqref{eq:rabi_theory}, \eqref{eq:rabi_eps}, and \eqref{eq:T2_geff}, versus the tunneling $\Gamma$ and its disorder $\sigma_\Gamma$ over the window of Fig.~\ref{fig:T2_ratio_rabi}, for $N=2$ (a-c) and $N=3$ (d-f) with the small gap (left column), large gap (middle column), and experimental (right column) parameters of Table~\ref{tab:parameters}.
    Each region is labeled by its shortest channel time, $T_{2,\text{Rabi},\Gamma}^{(N)}$, $T_{2,\text{Rabi},E_-}^{(N)}$, or $T_{2,\text{Rabi},\Gamma_{\text{eff}}}^{(N)}$, and the solid lines are the crossover boundaries of Eqs.~\eqref{eq:crossover_eps_geff}, \eqref{eq:crossover_gamma_geff}, and \eqref{eq:crossover_gamma_eps}.
    All other parameters are as in Table~\ref{tab:parameters}.}
    \label{fig:rabi_channels}
\end{figure*}

Combining the three dephasing channels of Eqs.~\eqref{eq:rabi_theory}, \eqref{eq:rabi_eps}, and \eqref{eq:T2_geff}, the phase of the Rabi oscillation is the sum of three random phases.
Approximating them as independent, their characteristic functions multiply, thus, in principle, the envelope of the oscillatory part of $\expval{P_{ee}(t)}$ is the product of the three envelopes,
\begin{equation}\label{eq:rabi_combined}
    \begin{split}
        \expval{P_{ee}(t)}\propto&\exp\left[-\frac{\sigma_\Gamma^2t^2}{2\hbar^2}\right]\left[1+\left(\frac{\sigma_{E_-}^2t}{\Gamma\hbar}\right)^2\right]^{-1/4}\\
        &\times
        \begin{cases}
            \left(1+x^2\right)^{-1/2}, & N=3,\\
            \left(1+x^2\right)^{-1}, & N=2,
        \end{cases}
    \end{split}
\end{equation}
with $x$ defined in Eq.~\eqref{eq:rabi_perp} and $\sigma_{E_-}^2$ of Eqs.~\eqref{eq:E_minus_variance_N3} and \eqref{eq:E_minus_variance_N2}, respectively.
However, for a useful heuristic estimate of the dominating Rabi dephasing channel, we can simply take the shortest of the three dephasing times,
\begin{equation}\label{eq:rabi_T2_min}
    T_{2,\text{Rabi}}^{(N)}\approx\min\left(T_{2,\text{Rabi},\Gamma}^{(N)},T_{2,\text{Rabi},E_-}^{(N)},T_{2,\text{Rabi},\Gamma_{\text{eff}}}^{(N)}\right).
\end{equation}
Figure~\ref{fig:rabi_channels} shows the estimate Eq.~\eqref{eq:rabi_T2_min} and the dominating channel over the $(\Gamma,\sigma_\Gamma)$ window of Fig.~\ref{fig:T2_ratio_rabi} for the three parameter sets of Table~\ref{tab:parameters} and both $N=2$ and $N=3$.

Here, the small gap set provides a good example to illustrate the crossover between all three dephasing channels.
In Fig.~\ref{fig:rabi_channels}(a), at small $\sigma_\Gamma$, the dominating channel crosses over from the second, $T_{2,\text{Rabi},E_-}^{(2)}$, to the third, $T_{2,\text{Rabi},\Gamma_{\text{eff}}}^{(2)}$, as $\Gamma$ increases, because the two have opposite trends with $\Gamma$: $T_{2,\text{Rabi},E_-}^{(N)}\propto\Gamma$ in Eq.~\eqref{eq:rabi_eps}, as a stronger drive suppresses the phase drift $E_-^2t/2\Gamma\hbar$, whereas $T_{2,\text{Rabi},\Gamma_{\text{eff}}}^{(N)}\propto1/\Gamma$ in Eq.~\eqref{eq:T2_geff}, as the fluctuation $\delta\Gamma_{\text{eff}}^{(N)}\propto\Gamma$ of Eqs.~\eqref{eq:gamma_eff_N3} and \eqref{eq:gamma_eff_N2} is amplified.
The crossover tunneling is estimated by equating Eqs.~\eqref{eq:rabi_eps} and \eqref{eq:T2_geff},
\begin{equation}\label{eq:crossover_eps_geff}
    \begin{split}
        \Gamma_{E_-\Gamma_{\text{eff}}}^{(3)}&=\frac{2\Delta}{\left(e^2+1\right)^{1/4}}\,\frac{\sigma_{E_-}(N=3)}{\sigma_\mu},\\
        \Gamma_{E_-\Gamma_{\text{eff}}}^{(2)}&=\frac{2\Delta}{\left[(e+1)(e^2+1)\right]^{1/4}}\,\frac{\sigma_{E_-}(N=2)}{\sigma_\mu},
    \end{split}
\end{equation}
which gives $\Gamma_{E_-\Gamma_{\text{eff}}}^{(2)}\approx1.8\ueV$ for the small gap set, the vertical boundary of the $E_-$ region in Fig.~\ref{fig:rabi_channels}(a), and $\Gamma_{E_-\Gamma_{\text{eff}}}^{(3)}\approx0.31\ueV$, below the window of Fig.~\ref{fig:rabi_channels}(d), since $\sigma_{E_-}(N=3)/\sigma_{E_-}(N=2)\sim\sigma_\mu/2\Delta$ by Eqs.~\eqref{eq:E_minus_variance_N2} and \eqref{eq:E_minus_variance_N3}; the second channel therefore never dominates for $N=3$ in the window.

The crossover boundary between the first and the third dephasing channels is estimated by equating Eqs.~\eqref{eq:rabi_theory} and \eqref{eq:T2_geff},
\begin{equation}\label{eq:crossover_gamma_geff}
    \begin{split}
        \Gamma_{\Gamma\Gamma_{\text{eff}}}^{(3)}&=\sqrt{\frac{e^2-1}{2}}\,\frac{(2\Delta)^2}{\sigma_\mu^2}\,\sigma_\Gamma,\\
        \Gamma_{\Gamma\Gamma_{\text{eff}}}^{(2)}&=\sqrt{\frac{e-1}{2}}\,\frac{(2\Delta)^2}{\sigma_\mu^2}\,\sigma_\Gamma,
    \end{split}
\end{equation}
which are straight lines through the origin of the $(\Gamma,\sigma_\Gamma)$ plane, the $N=3$ line being flatter by the factor $\sqrt{e+1}$ of Eq.~\eqref{eq:T2_geff_ratio}.
For the small gap set, they run from $(\Gamma,\sigma_\Gamma)\approx(1.8,0.030)\ueV$ to $(10,0.17)\ueV$ for $N=2$ [Fig.~\ref{fig:rabi_channels}(a)] and from $(2.3,0.02)\ueV$ to $(10,0.087)\ueV$ for $N=3$ [Fig.~\ref{fig:rabi_channels}(d)].
The $N=2$ line bounds the region where the third site can help at all, and accounts for the right branch of the U-shaped unity contour in Fig.~\ref{fig:T2_ratio_rabi}(a), which rises from $(\Gamma,\sigma_\Gamma)\approx(2.8,0.03)\ueV$ to $(10,0.17)\ueV$.
Below this line, at larger $\Gamma$ and small $\sigma_\Gamma$, $N=2$ is limited by $\Gamma_{\text{eff}}$ while $N=3$ is still limited by $\sigma_\Gamma$, so that the ratio $T_{2,\text{Rabi},\Gamma}^{(3)}/T_{2,\text{Rabi},\Gamma_{\text{eff}}}^{(2)}=\Gamma/\Gamma_{\Gamma\Gamma_{\text{eff}}}^{(2)}$ grows linearly from unity; below the $N=3$ line, both are limited by $\Gamma_{\text{eff}}$, and the improvement saturates at the fixed factor $\sqrt{e+1}\approx1.93$ of Eq.~\eqref{eq:T2_geff_ratio}, comparable to the factor $\sim 2$ reached along the lower edge $\sigma_\Gamma=0.02\ueV$ of Fig.~\ref{fig:T2_ratio_rabi}(a).

The crossover boundary between the first and the second dephasing channels is estimated by equating Eqs.~\eqref{eq:rabi_theory} and \eqref{eq:rabi_eps},
\begin{equation}\label{eq:crossover_gamma_eps}
    \Gamma_{\Gamma E_-}^{(N)}=\sqrt{\frac{2}{e^4-1}}\,\frac{\sigma_{E_-}^2}{\sigma_\Gamma},
\end{equation}
with $\sigma_{E_-}^2$ of Eq.~\eqref{eq:E_minus_variance_N2} or \eqref{eq:E_minus_variance_N3}, i.e., a hyperbola $\Gamma\sigma_\Gamma=\text{const}$.
For $N=2$ in the small gap parameter set, $\Gamma\sigma_\Gamma\approx0.054\ueV^2$, which bounds the $E_-$ region of Fig.~\ref{fig:rabi_channels}(a) from above, from $(\Gamma,\sigma_\Gamma)\approx(1,0.054)\ueV$ to the point $(1.8,0.030)\ueV$, and accounts for the left branch of the U in Fig.~\ref{fig:T2_ratio_rabi}(a), which descends from $(1.3,0.2)\ueV$ to $(2.8,0.03)\ueV$.
The actual branch lies well above the hyperbola because the algebraic envelope $[1+(\sigma_{E_-}^2t/\Gamma\hbar)^2]^{-1/4}$ drops faster than $\exp(-(\sigma_\Gamma^2t^2/2\hbar^2))$. 
Therefore, even though nominally $T_{2,\text{Rabi},\Gamma}^{(2)}$ is smallest, the envelope in Eq.~\eqref{eq:rabi_combined} is still largely affected by the second channel.
Below the hyperbola, at smaller $\Gamma$ and small $\sigma_\Gamma$, $N=2$ is dominated by the intrachain splitting, whereas for $N=3$, it remains dominated by $\sigma_\Gamma$ [Fig.~\ref{fig:rabi_channels}(d)].
The ratio is then $T_{2,\text{Rabi},\Gamma}^{(3)}/T_{2,\text{Rabi},E_-}^{(2)}=\Gamma_{\Gamma E_-}^{(2)}/\Gamma\propto1/(\Gamma\sigma_\Gamma)$.
Unlike the fixed factor $\sqrt{e+1}$ of the $\Gamma_{\text{eff}}$-limited region, it increases with smaller $\Gamma$ and $\sigma_\Gamma$ as long as $N=3$ remains limited by $\sigma_\Gamma$, and would saturate only once $T_{2,\text{Rabi},\Gamma}^{(3)}$ exceeds $T_{2,\text{Rabi},\Gamma_{\text{eff}}}^{(3)}$ or $T_{2,\text{Rabi},E_-}^{(3)}$, which occurs outside the plotting window [Fig.~\ref{fig:rabi_channels}(d)].

Above both boundaries, at larger $\sigma_\Gamma$, the Rabi coherence is simply limited by the first dephasing channel $T_{2,\text{Rabi},\Gamma}^{(N)}$ for both $N$, in which case the increase of the number of sites does not improve the coherence time; this is the interior of the U shape in Fig.~\ref{fig:T2_ratio_rabi}(a), where the fitted ratio is $\approx 1$.

For the large gap and experimental sets, the first dephasing channel dominates the entire window for both $N=2$ and $N=3$ [Fig.~\ref{fig:rabi_channels}(b,c,e,f)]: by Eqs.~\eqref{eq:crossover_gamma_geff} and \eqref{eq:crossover_gamma_eps}, the second and third dephasing channels of $N=2$ would dominate only for $\sigma_\Gamma\lesssim10^{-3}\ueV$ and $\sigma_\Gamma\lesssim10^{-2}\ueV$, respectively.
In Fig.~\ref{fig:T2_ratio_rabi}(c), we notice a weak improvement at small $\Gamma$, implying that a high-quality qubit manipulation requires not only a large gap and a longer chain but also a suppressed fluctuation of the junction coupling: the third site pays off only for $\sigma_\Gamma$ below the boundaries of Eqs.~\eqref{eq:crossover_gamma_geff} and \eqref{eq:crossover_gamma_eps}.

\subsection{Dephasing channels in Ramsey oscillation}\label{sec:ramsey_dephasing}
For the Ramsey oscillation, we consider the dephasing during the $\sigma_z$ rotation generated by $E_-$ at $\Gamma=0$, and ignore the dephasing in two $\pi/2$ $\sigma_x$ pulses since we assume zero fluctuation anyway as in Table~\ref{tab:parameters}.
Following the same logic as in Sec.~\ref{sec:interchain_tunneling} to derive Eq.~\eqref{eq:rabi_theory},  
the Ramsey coherence time is
\begin{equation}\label{eq:ramsey_T2}
    T_{2,\text{Ramsey}}^{(N)}=\frac{\sqrt2\hbar}{\sigma_{E_-}}.
\end{equation}
Therefore, it again boils down to the fluctuation of the splitting $E_-$, followed by the sum of single-chain energy splittings Eqs.~\eqref{eq:epsilon_N2} and \eqref{eq:epsilon_N3} for $N=2$ and $N=3$, respectively. 
Below we will show that, in $N=2$ the fluctuation $\sigma_{E_-}^2$ is not a function of $E_-$, whereas in $N=3$ it will increase with $E_-$, which is the main reason that the third site does not always improve the Ramsey coherence time.

For $N=2$, the energy splitting is 
\begin{equation}
    E_-(N=2)=u^{(2)}-\sum_{a\in\{L,R\}}\left(\nu_{a_1}+\frac{\mu_{a_1}\mu_{a_2}}{s_{a_1}}\right),
\end{equation}
and the variance $\sigma_{E_-}^2(N=2)$ is just Eq.~\eqref{eq:E_minus_variance_N2}.
For $N=3$, the energy splitting is 
\begin{widetext}
\begin{equation}
    \begin{split}
        &E_-(N=3) = \sum_{a\in\{L,R\}} - \frac{ \left( \nu_{a_1} - u^{(3)} \right) \left( \mu_{a_3} +\text{sgn}(E_-) u^{(3)} \right)}{2\Delta} - \frac{\mu_{a_1}\nu_{a_2}}{2\Delta} - \frac{\mu_{a_1}\mu_{a_2} \left( \mu_{a_3} +\text{sgn}(E_-) u^{(3)} \right)}{(2\Delta)^2} \\
        & = \sum_{a\in\{L,R\}} - \frac{\nu_{a_1} \mu_{a_3}}{2\Delta} - \frac{\mu_{a_1}\nu_{a_2}}{2\Delta} - \frac{\mu_{a_1}\mu_{a_2}\mu_{a_3}}{(2\Delta)^2} + u^{(3)} \frac{\mu_{a_3} - \text{sgn}(E_-) \nu_{a_1}}{2\Delta} - \text{sgn}(E_-) u^{(3)} \frac{\mu_{a_1}\mu_{a_2}}{(2\Delta)^2} + \text{sgn}(E_-) \frac{(u^{(3)})^2}{2\Delta},
    \end{split}
\end{equation}
\end{widetext}
Therefore, the variance is decomposed into two parts: the first part is the intrinsic splitting $\sigma_{E_-}^2(N=3)$ at the sweet spot independent of $u^{(3)}$ defined in Eq.~\eqref{eq:E_minus_variance_N3}, and the second part is the detuning-induced splitting linear in the magnitude of $u^{(3)}$ as
\begin{equation}
    \sigma_{E_-,u}^2(N=3)\approx\frac{\left(u^{(3)}\right)^2\left(\sigma_\nu^2+\sigma_\mu^2\right)}{2\Delta^2}\approx\frac{\abs{E_-}}{2\Delta}\sigma_\mu^2,
\end{equation}
where the first approximation ignores the higher-order term of $\mu^2$, and the
last approximation is because of Eq.~\eqref{eq:calibration} with $\abs{E_-}\ll \Delta$ and $\sigma_\nu\ll\sigma_\mu$.
Here, the crossover between $\sigma_{E_-}^2(N=3)$ and $\sigma_{E_-,u}^2(N=3)$ happens at $E_-=1.2\times10^{-2}$, $1.3\times10^{-4}$, and $1.1\times10^{-3}\ueV$ for the three sets. Therefore, the range in Fig.~\ref{fig:T2_ratio_ramsey} is dominated by the detuning-induced channel, except for the lowest decade of the small gap set, so we simply approximate $T_{2,\text{Ramsey}}^{(3)}\approx \frac{\sqrt{2}\hbar}{\sigma_{E_-,u}}\propto\abs{E_-}^{-1/2}$.
At $E_-=2\ueV$, this gives $T_{2,\text{Ramsey}}^{(3)}=1.07$, $19.1$, and $2.78$~ns for the three sets, consistent with Table~\ref{tab:comparison}.

Finally, the ratio of the two coherence times is
\begin{equation}\label{eq:ramsey_ratio}
    \frac{T_{2,\text{Ramsey}}^{(3)}}{T_{2,\text{Ramsey}}^{(2)}}=\frac{\sigma_{E_-}(N=2)}{\sigma_{E_-,u}(N=3)}\approx\sqrt{\frac{2\Delta\,\sigma_{E_-}^2(N=2)}{\sigma_\mu^2\abs{E_-}}}
\end{equation}
for $\abs{E_-}\ll\Delta$, which diverges as $\abs{E_-}^{-1/2}$ for $E_-\to0$. The crossover point is at $E_-=0.75$, $4.2$, and $0.46\ueV$ for the three parameter sets, qualitatively close to the numerical values $1.2$, $9.4$, and $1.4\ueV$ in Fig.~\ref{fig:T2_ratio_ramsey}(a-c). 
(where the discrepancy comes from other dephasing processes, e.g., two $\pi/2$ $\sigma_x$ pulses.)
Below the crossover, the power-law fits in Fig.~\ref{fig:T2_ratio_ramsey}(a-c) give the exponents $-0.40$, $-0.48$, and $-0.47$, verifying the $\abs{E_-}^{-1/2}$ divergence in Eq.~\eqref{eq:ramsey_ratio}; the smaller exponent of the small gap set is because its lowest decade lies below the crossover splitting $1.2\times10^{-2}\ueV$ of the intrinsic channel, where the ratio bends over toward the $E_-$-independent value $\sigma_{E_-}(N=2)/\sigma_{E_-}(N=3)$ set by the sweet-spot fluctuations alone.

Generally, the larger the energy splitting $E_-$ is, the shorter the coherence time of the three-site chain is, which is the consequence of the detuning away from the sweet spot; 
conversely, the improvement of up to two orders of magnitude at $E_-=10^{-3}\ueV$ is bought by a $\sigma_z$ rotation whose period $h/E_-$ grows faster than $T_{2,\text{Ramsey}}^{(3)}\propto\abs{E_-}^{-1/2}$, so the quality factor Eq.~\eqref{eq:quality} of the three-site chain, $Q\propto E_-T_{2,\text{Ramsey}}^{(3)}\propto\abs{E_-}^{1/2}$, still decreases toward small splittings, e.g., from $Q=62$ at $E_-=2\ueV$ to $Q=1.2$ at $E_-=10^{-3}\ueV$ for the large gap set, implying a trade-off between a good quantum memory and a good gate manipulation.

\begin{figure}[htbp]
    \centering
    \includegraphics[width=3.4in]{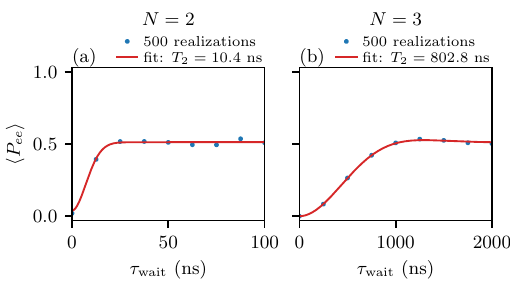}
    \caption{Failed Ramsey oscillation at a tiny splitting $E_-=10^{-3}\ueV$ for (a) $N=2$ and (b) $N=3$, with all other parameters being the large gap parameters in Table~\ref{tab:parameters}.
    The fitted $T_{2,\text{Ramsey}}$ of Eq.~\eqref{eq:fit} are $10.4$ and $802.8$~ns, respectively, both shorter than the nominal period $h/E_-\approx4.1\,\mu\text{s}$ of the oscillation.}
    \label{fig:ramsey_small_splitting}
\end{figure}

However, an energy splitting of $E_-=10^{-3}\ueV$, i.e., $E_-/k_B\approx12\,\mu\text{K}$, is not physically meaningful, as thermal and noise fluctuations would overwhelm it, and the associated quality factor is at most $Q=\mathcal{O}(1)$, implying that not even one complete Ramsey oscillation occurs within the coherence time, as shown in Fig.~\ref{fig:ramsey_small_splitting}.
Therefore, this regime is not relevant for any gate operation, but could in principle act as a quantum memory.

\section{Conclusion}\label{sec:conclusion}

In this work, inspired by a recent experimental breakthrough~\cite{zatelli2026majorana}, we have studied the effect of adding a pair of third sites on top of a double two-site quantum dot system in the Rabi and Ramsey oscillations~\cite{pan2025rabi}. 
We first derived the effective Hamiltonian of a general double $N$-site quantum dot system taking into account disorder effects in all microscopic parameters including the on-site chemical potential, normal intrachain tunneling, superconducting pairing, and the interchain tunneling.
This effective Hamiltonian allows us to induce the Rabi and Ramsey oscillations in a general double $N$-site quantum dot system.
We use the double three-site quantum dot system as an example to numerically simulate the Rabi and Ramsey oscillations in both the pristine limit and the disorder case in order to explicitly quantitatively estimate the coherence improvement in increasing the effective Majorana chain array size.
We find that the pristine Rabi oscillation is agnostic to the number of sites, giving the same oscillation return probability and leakage regardless of the number of sites.
However, the pristine Ramsey oscillation is sensitive to the number of sites: even though the return probability shows a similar oscillation, the leakage in the three-site case is significantly larger than in the two-site case because of the more microscopic parameters required to detune the system away from the sweet spot.
Beyond the pristine limit, we also study the actual Rabi and Ramsey oscillations in the presence of disorder for three different parameter sets, one of which is directly taken from the recent experiment~\cite{zatelli2026majorana}.
We find that the coherence time of the Rabi oscillation and Ramsey oscillation does not always increase with the number of sites: 
for the Rabi oscillation, it depends on the competition of three major dephasing channels, and if the noise is mainly in the interchain tunneling, the additional site can barely improve the coherence time.
The Ramsey improvement of the six-dot qubit only becomes large, up to two orders of magnitude, when the splitting is reduced to $\sim10^{-3}\ueV$, reinforcing the conclusion above regarding the trade-off between a good quantum memory and a good gate manipulation.
The $\sigma_z$ rotation generated by such a small splitting has a period of microseconds, longer than the coherence time itself, so the quality factor $Q$ decreases with the improvement.
In this limit, the qubit is tested as a quantum memory near the sweet spot rather than under gate manipulation.

This ``paradoxical'' result implies that using a longer chain only gives a better qubit in the sense of a better quantum memory, yet not necessarily a better qubit in terms of qubit gate operation. In experiments, we require not only a large gap and a longer chain to suppress the detuning from the sweet spot, but also a suppressed fluctuation of the interchain junction coupling because the additional sites only pay off for $\sigma_\Gamma$ below a certain threshold.

Alterative qubit manipulation protocols can be moved off the detuning: either by measurement-based operations~\cite{karzig2017scalable,tsintzis2024roadmap}, which replace the $\sigma_z$ rotation by parity measurements and never leave the sweet spot, or by a $\sigma_y$ rotation at the sweet spot, so that the full SU(2) single-qubit rotations can be generated by $\sigma_x$ and $\sigma_y$ without the leakage and the detuning-induced dephasing of the direct $\sigma_z$ rotation, which can be experimentally realized by second tunnel junction between the leftmost dots $L_1$ and $R_1$ of the two chains.

Our work demonstrates the difficulties in carrying out protected gate operations in the quantum dot Majorana platform~\cite{sau2012realizing} because even the fragile algebraic protection being compromised by the critical need to control far too many experimental parameters with increasing chain length.

\section*{Acknowledgments}
We thank C-X. Liu for the helpful comments.
H.P. is supported by startup funds at the University of Florida.
S.D.S. is supported by the Laboratory for Physical Sciences through the Condensed Matter Theory Center at the University of Maryland.
This work was partially performed at the Aspen Center for Physics, which is supported by National Science Foundation grant PHY-2210452.

\bibliography{Double_3_site_QD}
\appendix

\section{Splitting energy between the even and odd ground states}\label{app:splitting}
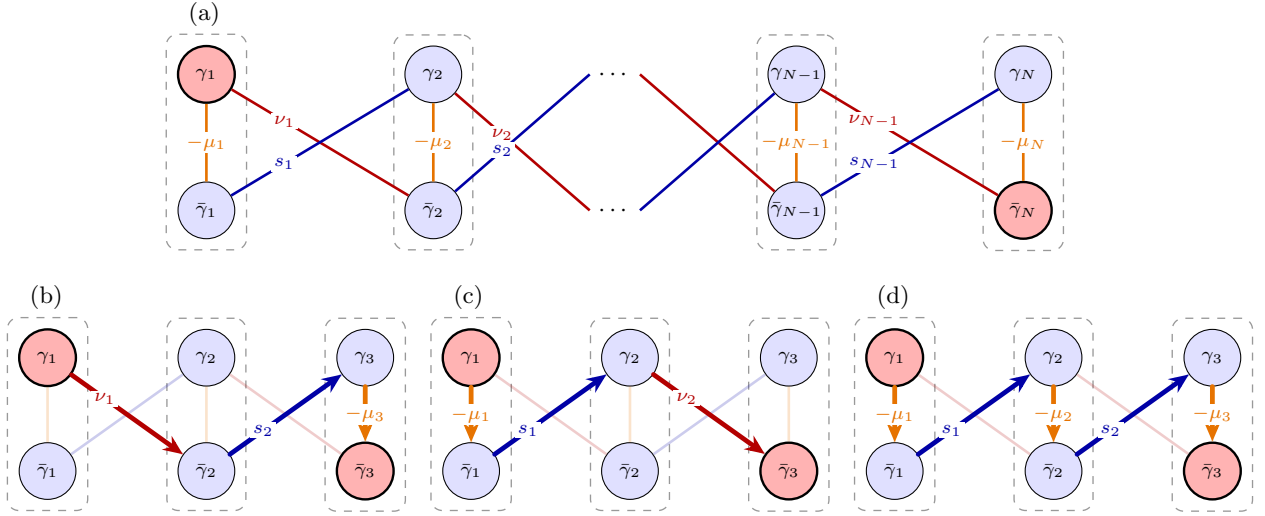
\begin{figure*}[htbp]
    \centering
    \begin{tikzpicture}[
        maj/.style={circle,draw=black,fill=blue!12,minimum size=7.5mm,inner sep=0pt,font=\scriptsize},
        endm/.style={maj,fill=red!30,line width=0.9pt},
        site/.style={rounded corners=1.5mm,draw=black!40,dashed,line width=0.5pt,inner sep=1.4mm},
        lmu/.style={line width=1pt,orange!90!black},
        lnu/.style={line width=1pt,red!70!black},
        ls/.style={line width=1pt,blue!65!black},
        faint/.style={line width=1pt,opacity=0.2},
        hop/.style={line width=1.8pt,line cap=round,-{Stealth[length=2.6mm,width=2.2mm]}},
        hmu/.style={hop,orange!90!black},
        hnu/.style={hop,red!70!black},
        hs/.style={hop,blue!65!black},
        lbl/.style={fill=white,inner sep=1pt,font=\scriptsize},
        panel/.style={font=\small,anchor=west}
    ]
        \begin{scope}[shift={(2.1,0)}]
            \node[panel] at (-0.35,1.7) {(a)};
            \node[endm] (g1) at (0,0.9) {$\gamma_{1}$};
            \node[maj]  (g2) at (3,0.9) {$\gamma_{2}$};
            \node[maj]  (g3) at (7.8,0.9) {$\gamma_{N-1}$};
            \node[maj]  (g4) at (10.8,0.9) {$\gamma_{N}$};
            \node[maj]  (b1) at (0,-0.9) {$\bar{\gamma}_{1}$};
            \node[maj]  (b2) at (3,-0.9) {$\bar{\gamma}_{2}$};
            \node[maj]  (b3) at (7.8,-0.9) {$\bar{\gamma}_{N-1}$};
            \node[endm] (b4) at (10.8,-0.9) {$\bar{\gamma}_{N}$};
            \node (dt) at (5.4,0.9) {$\cdots$};
            \node (db) at (5.4,-0.9) {$\cdots$};
            \node[site,fit=(g1)(b1)] {};
            \node[site,fit=(g2)(b2)] {};
            \node[site,fit=(g3)(b3)] {};
            \node[site,fit=(g4)(b4)] {};
            \draw[lmu] (g1) -- node[lbl] {$-\mu_{1}$} (b1);
            \draw[lmu] (g2) -- node[lbl] {$-\mu_{2}$} (b2);
            \draw[lmu] (g3) -- node[lbl] {$-\mu_{N-1}$} (b3);
            \draw[lmu] (g4) -- node[lbl] {$-\mu_{N}$} (b4);
            \draw[lnu] (g1) -- node[lbl,pos=0.3] {$\nu_{1}$} (b2);
            \draw[ls]  (b1) -- node[lbl,pos=0.3] {$s_{1}$} (g2);
            \draw[lnu] (g2) -- node[lbl,pos=0.35] {$\nu_{2}$} (db.west);
            \draw[ls]  (b2) -- node[lbl,pos=0.35] {$s_{2}$} (dt.west);
            \draw[lnu] (dt.east) -- (b3);
            \draw[ls]  (db.east) -- (g3);
            \draw[lnu] (g3) -- node[lbl,pos=0.3] {$\nu_{N-1}$} (b4);
            \draw[ls]  (b3) -- node[lbl,pos=0.3] {$s_{N-1}$} (g4);
        \end{scope}
        \majthree{0}{-3.6}{(b)}
        \draw[hnu] (g1) -- node[lbl,pos=0.3] {$\nu_{1}$} (b2);
        \draw[hs]  (b2) -- node[lbl,pos=0.3] {$s_{2}$} (g3);
        \draw[hmu] (g3) -- node[lbl] {$-\mu_{3}$} (b3);
        \majthree{5.6}{-3.6}{(c)}
        \draw[hmu] (g1) -- node[lbl] {$-\mu_{1}$} (b1);
        \draw[hs]  (b1) -- node[lbl,pos=0.3] {$s_{1}$} (g2);
        \draw[hnu] (g2) -- node[lbl,pos=0.3] {$\nu_{2}$} (b3);
        \majthree{11.2}{-3.6}{(d)}
        \draw[hmu] (g1) -- node[lbl] {$-\mu_{1}$} (b1);
        \draw[hs]  (b1) -- node[lbl,pos=0.3] {$s_{1}$} (g2);
        \draw[hmu] (g2) -- node[lbl] {$-\mu_{2}$} (b2);
        \draw[hs]  (b2) -- node[lbl,pos=0.3] {$s_{2}$} (g3);
        \draw[hmu] (g3) -- node[lbl] {$-\mu_{3}$} (b3);
    \end{tikzpicture}
    \caption{(a) Majorana coupling graph of Eq.~\eqref{eq:singlechain} for a single $N$-site chain, labeled by the elements of $K$ in Eq.~\eqref{eq:Klinks}: $-\mu_{i}$ between $\gamma_{i}$ and $\bar{\gamma}_{i}$, $\nu_{i}$ between $\gamma_{i}$ and $\bar{\gamma}_{i+1}$, and $s_{i}$ between $\bar{\gamma}_{i}$ and $\gamma_{i+1}$; dashed boxes are the dots. At the sweet spot only the bonds $s_{i}$ survive, leaving the end Majoranas $\gamma_{1}$ and $\bar{\gamma}_{N}$ (red) unpaired. (b)--(d) The three paths from $\gamma_{1}$ to $\bar{\gamma}_{3}$ for $N=3$, with the untraversed bonds faded. The product of the labels along a path, with $1/s_{j}$ for the strong bonds, is its contribution to $\kappa$ in Eq.~\eqref{eq:kappa}: $-\nu_{1}\mu_{3}/s_{2}$, $-\mu_{1}\nu_{2}/s_{1}$, and $-\mu_{1}\mu_{2}\mu_{3}/(s_{1}s_{2})$, the three terms of Eq.~\eqref{eq:epsilon_N3}.}
    \label{fig:paths}
\end{figure*}

In this section, we derive the energy splitting $\varepsilon$ between the even and odd ground states due to the detuning from the sweet spot, Eq.~\eqref{eq:epsilon}.
We rewrite Eq.~\eqref{eq:singlechain} as the Majorana bilinear form
\begin{equation}\label{eq:Kmatrix}
    \mathcal{H}=\frac{i}{4}\bm{\gamma}^\intercal K\bm{\gamma}+\sum_{i=1}^{N}\frac{\mu_{i}}{2},
\end{equation}
with $K$ real and antisymmetric, whose independent nonzero elements are the bonds of Eq.~\eqref{eq:epsilon},
\begin{equation}\label{eq:Klinks}
    K_{\gamma_{i},\bar{\gamma}_{i}}=-\mu_{i},\quad
    K_{\gamma_{i},\bar{\gamma}_{i+1}}=\nu_{i},\quad
    K_{\bar{\gamma}_{i},\gamma_{i+1}}=s_{i},
\end{equation}
as shown in Fig.~\ref{fig:paths}(a).
Alternatively, $iK$ is the single-particle Bogoliubov-de Gennes Hamiltonian for Eq.~\eqref{eq:singlechain} constructed in the Majorana basis.
We order the Majoranas as $\bm{\gamma}=(\bm{\gamma}_E,\bm{\gamma}_B)$, with the end sector $\bm{\gamma}_E=(\gamma_{1},\bar{\gamma}_{N})$ and the bulk sector $\bm{\gamma}_B=(\bar{\gamma}_{1},\gamma_{2},\bar{\gamma}_{2},\gamma_{3},\dots,\bar{\gamma}_{N-1},\gamma_{N})$ consisting of the $N-1$ dimers $(\bar{\gamma}_{j},\gamma_{j+1})$ paired by Eq.~\eqref{eq:Hsweet}, so that
\begin{equation}\label{eq:Kblocks}
    K=\begin{pmatrix}
        K_{EE} & K_{EB}\\
        K_{BE} & K_{BB}
    \end{pmatrix},\qquad K_{BE}=-K_{EB}^\intercal.
\end{equation}
For $N\geq3$ (the case $N=2$ can be directly diagonalized, as shown at the end of this appendix), the end Majoranas are not directly coupled, $K_{EE}=0$, and they couple to the bulk only through the weak bonds at the two ends,
\begin{equation}\label{eq:KEB}
    K_{EB}=\begin{pmatrix}
        -\mu_{1} & 0 & \nu_{1} & 0 & \cdots & 0\\
        0 & \cdots & 0 & -\nu_{N-1} & 0 & \mu_{N}
    \end{pmatrix}.
\end{equation}
The bulk block is $K_{BB}=K_S+K_W$, where
\begin{equation}\label{eq:KS}
    K_S=\bigoplus_{j=1}^{N-1}s_{j}J, \qquad J\equiv i\sigma_y
\end{equation}
collects the strong bonds, one $2\times2$ block per dimer, and $K_W$ the weak bonds $-\mu_{i}$ ($2\leq i\leq N-1$) and $\nu_{i}$ ($2\leq i\leq N-2$) between consecutive and next-to-consecutive dimers,
\begin{equation}\label{eq:KBB}
    K_{BB}={\setlength{\arraycolsep}{2.5pt}\begin{pmatrix}
        0 & s_{1} & 0 & 0 & 0 & 0 & 0 & \cdots\\
        -s_{1} & 0 & -\mu_{2} & 0 & \nu_{2} & 0 & 0 & \cdots\\
        0 & \mu_{2} & 0 & s_{2} & 0 & 0 & 0 & \cdots\\
        0 & 0 & -s_{2} & 0 & -\mu_{3} & 0 & \nu_{3} & \cdots\\
        0 & -\nu_{2} & 0 & \mu_{3} & 0 & s_{3} & 0 & \cdots\\
        0 & 0 & 0 & 0 & -s_{3} & 0 & -\mu_{4} & \cdots\\
        \vdots & \vdots & \vdots & \vdots & \vdots & \vdots & \vdots & \ddots
    \end{pmatrix}}.
\end{equation}

The $EE$ block of the Green's function corresponding to the single-particle Hamiltonian $iK$ is
\begin{equation}\label{eq:GEE}
    G_{EE}(\omega)^{-1}=\omega-iK_{EE}+K_{EB}(\omega-iK_{BB})^{-1}K_{BE}.
\end{equation}
Since the bulk is gapped and the end modes lie near zero energy, we approximate Eq.~\eqref{eq:GEE} by setting $\omega=0$ in the last term, $G_{EE}(\omega)^{-1}\approx\omega-iK_{\text{SE}}$, with the static self-energy of the end sector
\begin{equation}\label{eq:K_SE}
    K_{\text{SE}}=-K_{EB}K_{BB}^{-1}K_{BE}\equiv\kappa J,
\end{equation}
where the last equality holds because $K_{\text{SE}}$ is manifestly real and antisymmetric.
The many-body end-sector Hamiltonian is therefore
\begin{equation}\label{eq:H_SE}
    \mathcal{H}_{\text{SE}}=\frac{i}{4}\bm{\gamma}_E^\intercal K_{\text{SE}}\bm{\gamma}_E=\frac{\kappa}{2}\,i\gamma_{1}\bar{\gamma}_{N},
\end{equation}
up to parity-independent constants.

The sign relating $\kappa$ to $\varepsilon$ follows from the fermion parity $P=\prod_{i=1}^{N}(1-2n_{i})=\prod_{i=1}^{N}i\gamma_{i}\bar{\gamma}_{i}$.
We introduce the end mode $f$ and the bulk modes $d_{j}$ of the dimers $(\bar{\gamma}_{j},\gamma_{j+1})$,
\begin{equation}\label{eq:modes}
    f^\dagger=\frac{1}{2}\left[\gamma_{1}+i(-1)^{N-1}\bar{\gamma}_{N}\right],\quad
    d_{j}^\dagger=\frac{1}{2}\left(\gamma_{j+1}+i\bar{\gamma}_{j}\right),
\end{equation}
so that $i\gamma_{1}\bar{\gamma}_{N}=(-1)^{N-1}(1-2n_{f})$ and $i\bar{\gamma}_{j}\gamma_{j+1}=2n_{d_{j}}-1$, with $n_{f}=f^\dagger f$ and $n_{d_{j}}=d_{j}^\dagger d_{j}$; hence
\begin{equation}\label{eq:parity}
    \begin{split}
        P&=\left(i\gamma_{1}\bar{\gamma}_{N}\right)\prod_{j=1}^{N-1}\left(i\bar{\gamma}_{j}\gamma_{j+1}\right)\\
        &=\left(1-2n_{f}\right)\prod_{j=1}^{N-1}\left(1-2n_{d_{j}}\right).
    \end{split}
\end{equation}
Equation~\eqref{eq:Hsweet} reads $\mathcal{H}=\sum_{j=1}^{N-1}\Delta_{j}(2n_{d_{j}}-1)$, so its ground states are in the $d$ vacuum, on which $P=1-2n_{f}$: $\ket{e}$ is the $f$ vacuum and $\ket{o}$ the $f$-occupied state for all $N$.
Hence $i\gamma_{1}\bar{\gamma}_{N}=(-1)^{N-1}P$ equals $(-1)^{N-1}$ on $\ket{e}$ and $(-1)^{N}$ on $\ket{o}$, and Eq.~\eqref{eq:H_SE} gives
\begin{equation}\label{eq:eps_kappa}
    \varepsilon=E_e-E_o=(-1)^{N-1}\kappa.
\end{equation}

To evaluate $\kappa$, we expand $K_{BB}^{-1}$ in the weak bonds,
\begin{equation}\label{eq:neumann}
    K_{BB}^{-1}=\sum_{n=0}^{N-2}\left(-K_S^{-1}K_W\right)^{n}K_S^{-1}.
\end{equation}
Here the series terminates because $(K_S^{-1}K_W)^{N-1}=0$, and $K_S^{-1}$ remains off-diagonal and has the single element $(K_S^{-1})_{\bar{\gamma}_{j},\gamma_{j+1}}=-1/s_{j}$ out of the first Majorana of the dimer $j$, and $K_W$ has the two elements $(K_W)_{\gamma_{j+1},\bar{\gamma}_{j+1}}=-\mu_{j+1}$ and $(K_W)_{\gamma_{j+1},\bar{\gamma}_{j+2}}=\nu_{j+1}$ out of its second Majorana into the first Majoranas of the dimers $j+1$ and $j+2$, so $K_S^{-1}K_W$ connects $\bar{\gamma}_{j}$ only to $\bar{\gamma}_{j+1}$ and $\bar{\gamma}_{j+2}$ and, by antisymmetry, $\gamma_{j+1}$ only to $\gamma_{j}$ and $\gamma_{j-1}$.
Inserting Eq.~\eqref{eq:neumann} into Eq.~\eqref{eq:K_SE}, we find that the $n$th term of $\kappa=(K_{\text{SE}})_{12}$ is $-(-1)^{n}$ times the sum over bulk Majoranas $b_1,\dots,b_{2n+2}$ of
\begin{equation}\label{eq:walk}
    \begin{split}
        &(K_{EB})_{\gamma_{1}b_1}(K_S^{-1})_{b_1b_2}(K_W)_{b_2b_3}(K_S^{-1})_{b_3b_4}\cdots\\
        &\quad\cdots(K_W)_{b_{2n}b_{2n+1}}(K_S^{-1})_{b_{2n+1}b_{2n+2}}(K_{BE})_{b_{2n+2}\bar{\gamma}_{N}},
    \end{split}
\end{equation}
i.e., over the paths $P\in\mathcal{P}_N$ of Eq.~\eqref{eq:epsilon} with $n+1$ strong bonds, each traversed from $\bar{\gamma}_{j}$ to $\gamma_{j+1}$ and contributing $-1/s_{j}$, and $n+2$ weak bonds, each traversed from $\gamma_{i}$ and contributing the element $-\mu_{i}$ or $\nu_{i}$ of Eq.~\eqref{eq:Klinks}.
The signs $-(-1)^{n}(-1)^{n+1}=1$ cancel, so
\begin{equation}\label{eq:kappa}
    \begin{split}
        \kappa&=\sum_{P\in\mathcal{P}_N}\frac{\prod_{i\in P_\mu}(-\mu_{i})\prod_{j\in P_\nu}\nu_{j}}{\prod_{j\in P_s}s_{j}}\\
        &=(-1)^{N}\sum_{P\in\mathcal{P}_N}\frac{\prod_{i\in P_\mu}\mu_{i}\prod_{j\in P_\nu}\nu_{j}}{\prod_{j\in P_s}s_{j}},
    \end{split}
\end{equation}
where the second equality uses $\abs{P_\mu}=N-2\abs{P_\nu}$, as a path runs from left to right and covers every site exactly once, by the bond $\nu_{j}$ for the sites $j$ and $j+1$ or by the bond $\mu_{i}$ for the site $i$.
Together with Eq.~\eqref{eq:eps_kappa}, this is Eq.~\eqref{eq:epsilon}.

For $N=2$, $\gamma_{1}$ and $\bar{\gamma}_{2}$ are directly coupled, and the two entries $\pm\nu_{1}$ of Eq.~\eqref{eq:KEB} move to $K_{EE}=\nu_{1}J$, so that Eq.~\eqref{eq:GEE} gives $K_{\text{SE}}=K_{EE}-K_{EB}K_{BB}^{-1}K_{BE}$ in place of Eq.~\eqref{eq:K_SE}.
The term $K_{EE}$ supplies the direct path $\gamma_{1}\to\bar{\gamma}_{2}$ without strong bonds, and Eqs.~\eqref{eq:eps_kappa} and \eqref{eq:kappa} hold with this path included in $\mathcal{P}_{2}$.

Figure~\ref{fig:paths}(b)--(d) shows the three paths for $N=3$, as in Eq.~\eqref{eq:epsilon_N3}. For other $N$, for example, $N=2,\dots,5$, we can immediately write down the splitting energies as
\begin{equation}\label{eq:epsilon_enum}
    \begin{split}
        \varepsilon(N=2)&=-\nu_{1}-\frac{\mu_{1}\mu_{2}}{s_{1}},\\
        \varepsilon(N=3)&=-\frac{\nu_{1}\mu_{3}}{s_{2}}-\frac{\mu_{1}\nu_{2}}{s_{1}}-\frac{\mu_{1}\mu_{2}\mu_{3}}{s_{1}s_{2}},\\
        \varepsilon(N=4)&=-\frac{\nu_{1}\nu_{3}}{s_{2}}-\frac{\nu_{1}\mu_{3}\mu_{4}}{s_{2}s_{3}}-\frac{\mu_{1}\nu_{2}\mu_{4}}{s_{1}s_{3}}\\
        &\quad-\frac{\mu_{1}\mu_{2}\nu_{3}}{s_{1}s_{2}}-\frac{\mu_{1}\mu_{2}\mu_{3}\mu_{4}}{s_{1}s_{2}s_{3}},\\
        \varepsilon(N=5)&=-\frac{\nu_{1}\nu_{3}\mu_{5}}{s_{2}s_{4}}-\frac{\nu_{1}\mu_{3}\nu_{4}}{s_{2}s_{3}}-\frac{\mu_{1}\nu_{2}\nu_{4}}{s_{1}s_{3}}\\
        &\quad-\frac{\nu_{1}\mu_{3}\mu_{4}\mu_{5}}{s_{2}s_{3}s_{4}}-\frac{\mu_{1}\nu_{2}\mu_{4}\mu_{5}}{s_{1}s_{3}s_{4}}-\frac{\mu_{1}\mu_{2}\nu_{3}\mu_{5}}{s_{1}s_{2}s_{4}}\\
        &\quad-\frac{\mu_{1}\mu_{2}\mu_{3}\nu_{4}}{s_{1}s_{2}s_{3}}-\frac{\mu_{1}\mu_{2}\mu_{3}\mu_{4}\mu_{5}}{s_{1}s_{2}s_{3}s_{4}}.
    \end{split}
\end{equation}
A path has at least $\lceil N/2\rceil$ weak bonds, so for $\abs{\mu_{i}},\abs{\nu_{i}}\lesssim\delta\ll\Delta$ the splitting is of order $\delta^{\lceil N/2\rceil}/\Delta^{\lceil N/2\rceil-1}$ with $s_{j}\approx2\Delta$.

\section{Two-chain effective model}\label{app:effective_model}
In this section, we derive the effective Hamiltonian Eq.~\eqref{eq:effective} by integrating out the excited states.
The Green's function within the even-total-parity qubit subspace $\text{G}=\left\{ \ket{ee},\ket{oo} \right\}$ is
\begin{equation}\label{eq:green1}
    \mathcal{G}_{\text{GG}}(\omega)=\Pi\mathcal{G}(\omega)\Pi^\dagger,
\end{equation}
where $\Pi=\ketbra{ee}+\ketbra{oo}$ is the projection operator, and
\begin{equation}\label{eq:green}
    \begin{split}
        \mathcal{G}(\omega)&=\left(\omega-\mathcal{H}_{\text{tot}}\right)^{-1}\\
        &\equiv\begin{pmatrix}
            \omega-\mathcal{H}_{\text{GG}} & -\mathcal{H}_{\text{GX}}\\
            -\mathcal{H}_{\text{XG}} & \omega-\mathcal{H}_{\text{XX}}
        \end{pmatrix}^{-1},
    \end{split}
\end{equation}
where $\mathcal{H}_{\text{GG}}$ ($\mathcal{H}_{\text{XX}}$) is the ground-state (excited-state) sector of the total Hamiltonian $\mathcal{H}_{\text{tot}}$, and $\mathcal{H}_{\text{GX}}=\mathcal{H}_{\text{XG}}^\dagger$ is the overlap between the ground-state and excited-state sectors.

Inverting Eq.~\eqref{eq:green}, we find that the projected Green's function in Eq.~\eqref{eq:green1} is
\begin{equation}\label{eq:GGG}
    \mathcal{G}_{\text{GG}}(\omega)=\left[\omega-\mathcal{H}_{\text{GG}}-\Sigma(\omega)\right]^{-1},
\end{equation}
where the self-energy is
\begin{equation}\label{eq:selfenergy}
    \Sigma(\omega)=\mathcal{H}_{\text{GX}}\left(\omega-\mathcal{H}_{\text{XX}}\right)^{-1}\mathcal{H}_{\text{XG}}.
\end{equation}
Thus, the effective Hamiltonian defined on the even-total-parity qubit subspace $\left\{ \ket{ee},\ket{oo} \right\}$ is
\begin{equation}\label{eq:Heff_omega}
    H_{\text{eff}}(\omega)=\mathcal{H}_{\text{GG}}+\Sigma(\omega).
\end{equation}
Here, the first term $\mathcal{H}_{\text{GG}}$ is the total Hamiltonian $\mathcal{H}_{\text{tot}}=\mathcal{H}_L+\mathcal{H}_R+\mathcal{H}_{\text{tunn}}$ projected onto the even-total-parity ground-state subspace.
Since the tunneling term changes the parity of both chains, the left- and right-chain Hamiltonians give only the diagonal terms
\begin{equation}\label{eq:Hdiag}
    \begin{split}
        \mel{ee}{\mathcal{H}_{L}+\mathcal{H}_{R}}{ee}&=E_{e,L}+E_{e,R}=E_{ee},\\
        \mel{oo}{\mathcal{H}_{L}+\mathcal{H}_{R}}{oo}&=E_{o,L}+E_{o,R}=E_{oo},
    \end{split}
\end{equation}
and the tunneling term $\mathcal{H}_{\text{tunn}}$ gives the off-diagonal terms.
To evaluate the latter for general $N$, we write Eq.~\eqref{eq:Htunn} in terms of the Majorana operators of Sec.~\ref{sec:single_chain}, i.e., as the bond of Eq.~\eqref{eq:singlechain} between $L_N$ and $R_1$ with $t_{i}\to\Gamma$ and $\Delta_{i}\to0$,
\begin{equation}\label{eq:Htunn_majorana}
    \mathcal{H}_{\text{tunn}}=\frac{i\Gamma}{2}\bar{\gamma}_{L_N}\gamma_{R_1}-\frac{i\Gamma}{2}\gamma_{L_N}\bar{\gamma}_{R_1}.
\end{equation}
At the sweet spot Eq.~\eqref{eq:sweetspot}, $\bar{\gamma}_{L_N}$ and $\gamma_{R_1}$ are the end Majoranas of the two chains adjacent to the junction, whereas $\gamma_{L_N}$ and $\bar{\gamma}_{R_1}$ belong to the dimers $(\bar{\gamma}_{L_{N-1}},\gamma_{L_N})$ and $(\bar{\gamma}_{R_1},\gamma_{R_2})$ paired by Eq.~\eqref{eq:Hsweet} and each creates a bulk excitation, so that the second term of Eq.~\eqref{eq:Htunn_majorana} has no matrix element within the ground-state subspace.
The end Majoranas act on the two product states $\ket{\pm}\equiv\prod_{i=1}^{N}P_{i}^{\pm}\ket{0}$ of Eq.~\eqref{eq:eo_states} as
\begin{equation}\label{eq:end_action}
    \gamma_{1}\ket{\pm}=\pm\ket{\pm},\qquad
    \bar{\gamma}_{N}\ket{\pm}=\pm i(-1)^{N-1}\ket{\mp},
\end{equation}
since $\gamma_{1}=c_{1}+c_{1}^{\dagger}$ acts on the leftmost factor, $\gamma_{1}P_{1}^{\pm}\ket{\chi}=\pm P_{1}^{\pm}\ket{\chi}$ for any $\ket{\chi}$ with site $1$ empty, whereas $\bar{\gamma}_{N}=i(c_{N}-c_{N}^{\dagger})$ anticommutes with $c_{i}^{\dagger}$ and thus turns each $P_{i}^{\pm}$ with $i<N$ into $P_{i}^{\mp}$ before acting on the rightmost one, $\bar{\gamma}_{N}P_{N}^{\pm}\ket{0}=\pm i(-1)^{N-1}P_{N}^{\mp}\ket{0}$.
Hence $\gamma_{1}\ket{o}=\ket{e}$ and $\bar{\gamma}_{N}\ket{o}=i(-1)^{N-1}\ket{e}$ in each chain.
For $\ket{oo}=\ket{o}_L\ket{o}_R$, with the left-chain operators ordered first, $\gamma_{R_1}$ acquires a sign $-1$ on passing the odd state $\ket{o}_L$, so $\bar{\gamma}_{L_N}\gamma_{R_1}\ket{oo}=-i(-1)^{N-1}\ket{ee}$ and
\begin{equation}\label{eq:Hoffdiag}
    \mel{ee}{\mathcal{H}_{\text{tunn}}}{oo}=\mel{oo}{\mathcal{H}_{\text{tunn}}}{ee}=(-1)^{N-1}\frac{\Gamma}{2}.
\end{equation}
The sign $(-1)^{N-1}$ is removed by the redefinition $\ket{oo}\to(-1)^{N-1}\ket{oo}$, which leaves $\sigma_z$ invariant and which we adopt throughout.

Therefore, the first term in the effective Hamiltonian is, in the basis $\left\{ \ket{ee},\ket{oo} \right\}$,
\begin{equation}\label{eq:HGG}
    H=\begin{pmatrix}
        E_{ee} & \Gamma/2\\
        \Gamma/2 & E_{oo}
    \end{pmatrix}
    =\frac{E_{+}}{2}\mathds{1}+\frac{E_{-}}{2}\sigma_z+\frac{\Gamma}{2}\sigma_x,
\end{equation}
where $\sigma_z=\ketbra{ee}-\ketbra{oo}$ and $\sigma_x=\ketbra{ee}{oo}+\ketbra{oo}{ee}$, which is Eq.~\eqref{eq:effective}.

The second term, the self-energy $\Sigma(\omega)$, accounts for the excited-state energies and the leakage from the ground-state subspace (computational basis) to the excited states (non-computational basis).
The coupling contributes to a scalar energy shift at the strength of $\mathcal{O}(\Gamma^{2})$.
Therefore, the self-energy term $\Sigma(\omega)$ can be neglected in the weak-tunneling limit $\Gamma\ll\Delta$, and the effective Hamiltonian is simply Eq.~\eqref{eq:HGG}.

\section{Analytical expression for $P_{ee}$ and $P_{\text{leak}}$ for the Rabi oscillation in the double $N$-site quantum-dot chain}\label{app:rabi_analytical}
In this section, we consider a general double-chain system with $N$ sites to derive Eqs.~\eqref{eq:P_ee_rabi} and \eqref{eq:P_leak_rabi} in Sec.~\ref{sec:rabi}, and to show that the Rabi oscillation is agnostic to the number of sites $N$ in the chain in the pristine limit, namely, both the oscillation frequency and the leakage to the excited states are independent of $N$.

At the sweet spot Eq.~\eqref{eq:sweetspot}, the single-chain Hamiltonian Eq.~\eqref{eq:Hsweet} is diagonal in the bulk modes $d_{i}$ of Eq.~\eqref{eq:modes},
\begin{equation}\label{eq:bulk_modes}
    \mathcal{H}=\sum_{i=1}^{N-1}\left(-\Delta_{i}+2\Delta_{i}d_{i}^\dagger d_{i}\right),
\end{equation}
whose two ground states are the vacuum of all $d_{i}$, labeled by the occupation of the end mode $f$ of Eq.~\eqref{eq:modes} according to Eq.~\eqref{eq:parity}, i.e.,
\begin{equation}\label{eq:GS}
    \ket{e,o}=\left( \bigotimes_{i=1}^{N-1} \ket{n_{d_i}=0} \right)\ket{n_{f}=0,1} ,
\end{equation}  
with $f^\dagger\ket{e}=\ket{o}$ for all $N$ [Eq.~\eqref{eq:end_action}].

From Eq.~\eqref{eq:Htunn_majorana}, we can separate the tunneling term into bulk and end modes,
\begin{equation}\label{eq:tunnel_majorana}
        \mathcal{H}_{\text{tunn}}=\underbrace{-\frac{i\Gamma}{2}\gamma_{L_N}\bar{\gamma}_{R_1}}_{\text{bulk}}+\underbrace{\frac{i\Gamma}{2}\bar{\gamma}_{L_N}\gamma_{R_1}}_{\text{end}}\equiv\mathcal{H}_{\text{tunn}}^{B}+\mathcal{H}_{\text{tunn}}^{E}.
\end{equation}
Therefore, starting from the ground state $\ket{ee}$, the end term $\mathcal{H}_{\text{tunn}}^{E}=\frac{i\Gamma}{2}\bar{\gamma}_{L_N}\gamma_{R_1}$ will mix $\ket{ee}$ and the ground state in the other parity $\ket{oo}$, while the bulk term $\mathcal{H}_{\text{tunn}}^{B}=-\frac{i\Gamma}{2}\gamma_{L_N}\bar{\gamma}_{R_1}$ will bring the bulk ground state $\ket{00}_d$ into the excited state $d^\dagger_{L_{N-1}}d^\dagger_{R_1}\ket{00}_d\equiv\ket{11}_d$, where $\ket{00}_d$ is the vacuum of the bulk modes with all $n_{d_{L_i}}=0$ and $n_{d_{R_i}}=0$ defined in Eq.~\eqref{eq:GS} for both left and right chains.
Namely, there are only four possible states to land starting from $\ket{ee}$, whose Hilbert space is factored as
\begin{equation}\label{eq:rabi_basis}
    B\otimes E=\left\{\ket{00}_d,\ket{11}_d\right\}\otimes\left\{\ket{00}_f,\ket{11}_f\right\},
\end{equation}
where $f$ is the end mode which flips the parity of the single chain.
Here and below, the two digits of a ket with the mode subscript $d$ ($f$) are the occupations of that mode on the left and on the right chain, in this order, i.e., $\ket{n_{d_{L_{N-1}}}n_{d_{R_1}}}_{d}$ ($\ket{n_{f_L}n_{f_R}}_{f}$); they are not the single-chain kets $\ket{n_{f}n_{d}}$ of Appendices~\ref{app:ramsey_N3} and \ref{app:ramsey_analytical}.

To be explicit, the many-body wave functions for the four states are
\begin{equation}\label{eq:rabi_states}
    \begin{split}
        \ket{ee}&=\ket{n_{d_{L_{N-1}}}=0,n_{f_L}=0,n_{d_{R_1}}=0,n_{f_R}=0},\\
        \ket{oo}&=\ket{n_{d_{L_{N-1}}}=0,n_{f_L}=1,n_{d_{R_1}}=0,n_{f_R}=1},\\
        \ket{o'o'}&=\ket{n_{d_{L_{N-1}}}=1,n_{f_L}=0,n_{d_{R_1}}=1,n_{f_R}=0},\\
        \ket{e'e'}&=\ket{n_{d_{L_{N-1}}}=1,n_{f_L}=1,n_{d_{R_1}}=1,n_{f_R}=1},
    \end{split}
\end{equation}
where $\ket{e'}$ ($\ket{o'}$) denotes the excited state in the even (odd) parity sector, and all other bulk modes are in the vacuum state, i.e., $n_{d_{L_i}}=0$ for $i=1,\dots,N-2$ and $n_{d_{R_i}}=0$ for $i=2,\dots,N-1$ (i.e., Eq.~\eqref{eq:GS}).

Since the two terms in Eq.~\eqref{eq:tunnel_majorana} act on disjoint sets of modes, they manifestly commute, and the evolution operator factorizes simply as
\begin{equation}\label{eq:U_factor}
    \begin{split}
        U(\tau)&=e^{-i\mathcal{H}_{\text{tot}}\tau/\hbar}=e^{-i\left(\mathcal{H}_{B}\otimes\mathds{1}_E+\mathds{1}_B\otimes\mathcal{H}_E\right)\tau/\hbar}\\
        &=e^{-i\mathcal{H}_{B}\tau/\hbar}e^{-i\mathcal{H}_E\tau/\hbar}=U_{B}(\tau)\otimes U_{E}(\tau),
    \end{split}
\end{equation}
where we define $\mathcal{H}_{B}\equiv\mathcal{H}_L+\mathcal{H}_R+\mathcal{H}_{\text{tunn}}^{B}$ spanned only by the bulk modes and $\mathcal{H}_{E}\equiv\mathcal{H}_{\text{tunn}}^{E}$ spanned only by the end modes, which commute because $\bar{\gamma}_{L_N}$ and $\gamma_{R_1}$ do not appear in Eq.~\eqref{eq:Hsweet}.
Here,
\begin{equation}\label{eq:U_B}
    U_B(\tau)=\exp\left[-\frac{i\tau}{\hbar}\begin{pmatrix}
        E_{G} & -\Gamma/2\\
        -\Gamma/2 & E_{X}
    \end{pmatrix}\right],
\end{equation}
where the diagonal elements are
\begin{equation}\label{eq:E_GX}
    \begin{split}
        \bra{00}_d\mathcal{H}_{B}\ket{00}_d&\equiv E_G=-2(N-1)\Delta,\\
        \bra{11}_d\mathcal{H}_{B}\ket{11}_d&\equiv E_X=-2(N-3)\Delta,
    \end{split}
\end{equation}
with the bulk excited energy $E_X$ due to the excited gap in each chain being $2\Delta$, and the off-diagonal is
\begin{equation}\label{eq:HB_offdiag}
    \bra{00}_d\mathcal{H}_{B}\ket{11}_d=\bra{11}_d\mathcal{H}_{B}\ket{00}_d=-\Gamma/2.
\end{equation}

For the edge mode evolution, we have
\begin{equation}\label{eq:U_E}
    U_E(\tau)=\exp\left[-\frac{i\tau}{\hbar}\begin{pmatrix}
        0 & \Gamma/2\\
        \Gamma/2 & 0
    \end{pmatrix}\right],
\end{equation}
given that only the off-diagonal elements are finite [Eq.~\eqref{eq:Hoffdiag}]
\begin{equation}\label{eq:HE_offdiag}
    \bra{00}_f\mathcal{H}_E\ket{11}_f=\bra{11}_f\mathcal{H}_E\ket{00}_f=\Gamma/2.
\end{equation}

Therefore, with the initial state $\ket{ee}=(1,0)^\intercal\otimes(1,0)^\intercal$, the instantaneous state at time $\tau$ in the Rabi oscillations is, up to a global phase,
\begin{equation}\label{eq:psi_rabi}
    \begin{split}
        \ket{\psi(\tau)}&=U(\tau)\ket{ee}=U_B(\tau)\ket{00}_d\otimes U_E(\tau)\ket{00}_f\\
        &=\begin{pmatrix}
            \cos\frac{\Omega\tau}{\hbar}+i\sin\frac{\Omega\tau}{\hbar}\frac{2\Delta}{\Omega}\\
            i\sin\frac{\Omega\tau}{\hbar}\frac{\Gamma}{2\Omega}
        \end{pmatrix}\otimes\begin{pmatrix}
            \cos\frac{\Gamma\tau}{2\hbar}\\
            -i\sin\frac{\Gamma\tau}{2\hbar}
        \end{pmatrix},
    \end{split}
\end{equation}
where $\Omega=\sqrt{\frac{\Gamma^2}{4}+4\Delta^2}$.

Therefore the leakage probability to the bulk excited states is Eq.~\eqref{eq:P_leak_rabi}, and within the computational basis, the probabilities of finding $\ket{ee}$ and $\ket{oo}$ in a Rabi oscillation are Eq.~\eqref{eq:P_ee_rabi} and
\begin{equation}\label{eq:P_oo_rabi}
    P_{oo}(\tau)=\sin^2\left(\frac{\Gamma}{2}\tau/\hbar\right)\left(1-P_{\text{leak}}(\tau)\right),
\end{equation}
respectively.

\section{Ramsey oscillation in the double three-site quantum-dot chain}\label{app:ramsey_N3}
In this section, we derive the three-site chain splitting $E_-$ which can be inversely solved to derive the detuning amount $u^{(3)}$ in Eq.~\eqref{eq:calibration} of Sec.~\ref{sec:ramsey}.

For $E_->0$, Eq.~\eqref{eq:ramsey_detuning_N3} sets $\nu_{1}=-u^{(3)}$, $\mu_{3}=u^{(3)}$, $s_{1}=2\Delta+u^{(3)}$, and $s_{2}=2\Delta$, with $\mu_{1}=\mu_{2}=\nu_{2}=0$ in Eq.~\eqref{eq:singlechain} (see the path of couplings in Fig.~\ref{fig:paths}(b)).
The pair $(\bar{\gamma}_{1},\gamma_{2})$ is coupled only through $s_{1}$, and is decoupled from the remaining four Majoranas $(\gamma_{1},\bar{\gamma}_{2},\gamma_{3},\bar{\gamma}_{3})$ connected by $\nu_{1}$, $s_{2}$, and $\mu_{3}$ along the path of Fig.~\ref{fig:paths}(b).
Therefore, we can focus on an effective chain of the four Majoranas $(\gamma_{1},\bar{\gamma}_{2},\gamma_{3},\bar{\gamma}_{3})$.
In terms of the end mode $f^\dagger=\frac{1}{2}(\gamma_{1}+i\bar{\gamma}_{3})$ and the bulk mode $d\equiv d_{2}$, $d^\dagger=\frac{1}{2}(\gamma_{3}+i\bar{\gamma}_{2})$, of the dimer $(\bar{\gamma}_{2},\gamma_{3})$ [Eq.~\eqref{eq:modes} with $N=3$], Eq.~\eqref{eq:singlechain} simplifies to (up to a parity-independent constant)
\begin{equation}\label{eq:H_wait}
    \mathcal{H}=\Delta\left(2d^\dagger d-1\right)+u^{(3)}\left(f^\dagger d+d^\dagger f\right),
\end{equation}
which conserves $n_f+n_d$.
In the occupation basis $\ket{n_{f}n_{d}}$ of the single chain, each parity sector therefore reduces to a $2\times2$ block,
\begin{equation}\label{eq:sector_blocks}
    \mathcal{H}^{e}=\begin{pmatrix}
        -\Delta & 0 \\
        0 & \Delta
    \end{pmatrix}_{\{\ket{00},\ket{11}\}},
    \quad
    \mathcal{H}^{o}=\begin{pmatrix}
        -\Delta & u^{(3)} \\
        u^{(3)} & \Delta
    \end{pmatrix}_{\{\ket{10},\ket{01}\}},
\end{equation}
with $\ket{e}=\ket{00}$ and $\ket{o}=\ket{10}$ [cf. Eq.~\eqref{eq:parity}].
The two digits of $\ket{n_{f}n_{d}}$, here and in Appendix~\ref{app:ramsey_analytical}, are the end-mode and bulk-mode occupations of one chain, in this order; they are not the two-chain kets $\ket{00}_{d}$, $\ket{11}_{d}$, $\ket{00}_{f}$, and $\ket{11}_{f}$ of Appendix~\ref{app:rabi_analytical}, whose digits are the left- and right-chain occupations of one mode.
Therefore, the eigenenergies of chain $a$ in the even and odd parity sectors are $\pm\Delta$ and $\pm\sqrt{\Delta^2+(u^{(3)})^2}$, respectively, with the sector ground-state energies
\begin{equation}
    E_{e,a}=-\Delta, \quad E_{o,a}=-\sqrt{\Delta^2+(u^{(3)})^2},
\end{equation}
leading to the exact splitting
\begin{equation}
    \varepsilon_a=E_{e,a}-E_{o,a}=\sqrt{\Delta^2+(u^{(3)})^2}-\Delta.
\end{equation}
Therefore, the total splitting of the two chains is
\begin{equation}\label{eq:eps_exact}
    E_-=E_{ee}-E_{oo}=2\left(\sqrt{\Delta^2+(u^{(3)})^2}-\Delta\right),
\end{equation}
whose leading order is just $(u^{(3)})^2/\Delta$.
Inverting Eq.~\eqref{eq:eps_exact} for $u^{(3)}$ gives the calibration Eq.~\eqref{eq:calibration}.
Similarly, for $E_-<0$, the sign reversal of $\mu_{a_3}$ in Eq.~\eqref{eq:ramsey_detuning_N3} exchanges the roles of the even and odd sectors in Eq.~\eqref{eq:sector_blocks}, so that Eq.~\eqref{eq:eps_exact} holds with $E_-$ replaced by $\abs{E_-}$ and Eq.~\eqref{eq:calibration} follows for both signs of $E_-$.

\section{Analytical expression for $P_{ee}$ and $P_{\text{leak}}$ for the Ramsey oscillation in the double three-site quantum-dot chain}\label{app:ramsey_analytical}
In this section, we derive Eqs.~\eqref{eq:P_leak_ramsey_main} and \eqref{eq:P_ee_ramsey_main} of Sec.~\ref{sec:ramsey}.
The two parity blocks of Eq.~\eqref{eq:sector_blocks} give the leakage during the wait of the Ramsey sequence in closed form.
For $E_->0$, since $\ket{e}=\ket{00}$ is an eigenstate of $\mathcal{H}^{e}$, only $\ket{o}=\ket{10}$ leaks to the bulk excited state $\ket{01}$, in the single-chain basis $\ket{n_{f}n_{d}}$ of Appendix~\ref{app:ramsey_N3}; below, the subscript $a\in\{L,R\}$ of $\ket{n_{f}n_{d}}_{a}$ labels the chain.

We treat the two $\pi/2$ pulses as ideal logical rotations $e^{-i\pi\sigma_x/4}=\left(\mathds{1}-i\sigma_x\right)/\sqrt{2}$ [Eq.~\eqref{eq:effective} at $E_-=0$].
The pulses are not leakage free: by the factorization Eq.~\eqref{eq:U_factor}, a pulse maps $\ket{00}_d\otimes\ket{\ell}$ to $U_B(\tau_{\pi/2})\ket{00}_d\otimes e^{-i\pi\sigma_x/4}\ket{\ell}$ for any logical state $\ket{\ell}$, with $\ket{00}_d$ the two-chain bulk vacuum of Eq.~\eqref{eq:rabi_basis} rather than a single-chain ket [Eqs.~\eqref{eq:U_E} and \eqref{eq:psi_rabi}], so that the logical rotation is exact but the common bulk factor leaks with the probability Eq.~\eqref{eq:P_leak_rabi} at $\tau=\tau_{\pi/2}$.
This leakage is second order in $\Gamma_{\pi/2}/\Delta$ and is the only leakage of the $N=2$ sequence [Fig.~\ref{fig:pristine_ramsey}(c)]; we neglect it relative to the leakage during the wait, which is first order in $\abs{E_-}/\Delta$ as shown below.
The first pulse then prepares
\begin{equation}\label{eq:psi1_ramsey}
    \ket{\psi_{1}}=\frac{1}{\sqrt{2}}\left(\ket{ee}-i\ket{oo}\right).
\end{equation}
During the wait (the $\sigma_z$ rotation), the $\ket{ee}=\ket{00}_L\ket{00}_R$ branch only picks up the phase $e^{2i\Delta\tau/\hbar}$ from $\mathcal{H}^{e}$, whereas each chain of the $\ket{oo}=\ket{10}_L\ket{10}_R$ branch evolves under $\mathcal{H}^{o}$ with
\begin{equation}\label{eq:U_wait}
    \begin{split}
        \mathcal{U}_{\tau}&\equiv e^{-i\mathcal{H}^{o}\tau/\hbar}\\
        &=\begin{pmatrix}
            \cos\phi+ik\sin\phi & -i\sqrt{1-k^{2}}\sin\phi\\
            -i\sqrt{1-k^{2}}\sin\phi & \cos\phi-ik\sin\phi
        \end{pmatrix}_{\{\ket{10},\ket{01}\}},
    \end{split}
\end{equation}
where $\phi\equiv\sqrt{\Delta^{2}+(u^{(3)})^{2}}\,\tau/\hbar=\left(2\Delta+\abs{E_-}\right)\tau/2\hbar$ and $k\equiv\Delta/\sqrt{\Delta^{2}+(u^{(3)})^{2}}=2\Delta/\left(2\Delta+\abs{E_-}\right)$ by Eq.~\eqref{eq:eps_exact}, so that $\sqrt{1-k^{2}}=u^{(3)}/\sqrt{\Delta^{2}+(u^{(3)})^{2}}$.
Namely, $[\mathcal{U}_{\tau}]_{11}=\mel{10}{\mathcal{U}_{\tau}}{10}$ is the logical amplitude and $[\mathcal{U}_{\tau}]_{21}=\mel{01}{\mathcal{U}_{\tau}}{10}$ the leakage amplitude of one chain, and Eq.~\eqref{eq:psi1_ramsey} evolves into
\begin{equation}\label{eq:psi2_ramsey}
    \begin{split}
        \ket{\psi_{2}(\tau)}=\frac{1}{\sqrt{2}}\Big\{&e^{2i\Delta\tau/\hbar}\ket{ee}-i[\mathcal{U}_{\tau}]_{11}^{2}\ket{oo}\\
        &-i[\mathcal{U}_{\tau}]_{11}[\mathcal{U}_{\tau}]_{21}\left(\ket{10}_L\ket{01}_R+\ket{01}_L\ket{10}_R\right)\\
        &-i[\mathcal{U}_{\tau}]_{21}^{2}\ket{01}_L\ket{01}_R\Big\},
    \end{split}
\end{equation}
where the last two lines are the components in which one or both chains have leaked to the bulk, with the total weight $\frac{1}{2}\left(1-\abs{[\mathcal{U}_{\tau}]_{11}}^{4}\right)$.

The second pulse acts on the logical part of Eq.~\eqref{eq:psi2_ramsey} as $\left(\mathds{1}-i\sigma_x\right)/\sqrt{2}$, while the leaked part stays orthogonal to the logical subspace: at the sweet spot, $\mathcal{H}_{\text{tunn}}$ of Eq.~\eqref{eq:tunnel_majorana} only flips $n_{f_L}$ and $n_{f_R}$ together or $n_{d_{L_2}}$ and $n_{d_{R_1}}$ together, so that $n_{d_{R_2}}$ and $(-1)^{n_{d_{L_2}}+n_{d_{R_1}}}$ are conserved during the pulse, and every leaked component of Eq.~\eqref{eq:psi2_ramsey} has $n_{d_{R_2}}=1$ or $n_{d_{L_2}}+n_{d_{R_1}}=1$, in contrast to the logical states with $n_{d_{L_2}}=n_{d_{R_1}}=n_{d_{R_2}}=0$.
Hence, up to a global phase,
\begin{equation}\label{eq:psi3_ramsey}
    \begin{split}
        \ket{\psi_{3}}=\frac{1}{2}\Big\{&\left[1-\left(e^{-i\Delta\tau/\hbar}[\mathcal{U}_{\tau}]_{11}\right)^{2}\right]\ket{ee}\\
        &-i\left[1+\left(e^{-i\Delta\tau/\hbar}[\mathcal{U}_{\tau}]_{11}\right)^{2}\right]\ket{oo}\Big\}+\ket{\psi_{\text{leak}}},
    \end{split}
\end{equation}
where $\ket{\psi_{\text{leak}}}$ is the image of the leaked part of Eq.~\eqref{eq:psi2_ramsey} under the second pulse and has the same weight.
Only the ratio of the two logical amplitudes of Eq.~\eqref{eq:psi2_ramsey} enters, which we write as
\begin{equation}\label{eq:R_ramsey}
    \begin{split}
        R(\tau)&\equiv e^{-i\Delta\tau/\hbar}[\mathcal{U}_{\tau}]_{11}=e^{-i\Delta\tau/\hbar}\left(\cos\phi+ik\sin\phi\right)\\
        &\equiv\sqrt{1-P_{01}(\tau)}\,e^{i\chi(\tau)},
    \end{split}
\end{equation}
whose modulus is fixed by the single-chain leakage probability
\begin{equation}\label{eq:P01}
    \begin{split}
        P_{01}(\tau)&=\abs{[\mathcal{U}_{\tau}]_{21}}^{2}=\left(1-k^{2}\right)\sin^{2}\phi\\
        &=\frac{\abs{E_-}\left(4\Delta+\abs{E_-}\right)}{\left(2\Delta+\abs{E_-}\right)^{2}}\sin^{2}\left(\frac{\left(2\Delta+\abs{E_-}\right)\tau}{2\hbar}\right),
    \end{split}
\end{equation}
since $\abs{[\mathcal{U}_{\tau}]_{11}}^{2}+\abs{[\mathcal{U}_{\tau}]_{21}}^{2}=1$.
Equation~\eqref{eq:psi3_ramsey} gives $P_{ee}=\abs{1-R^{2}}^{2}/4$ and $P_{oo}=\abs{1+R^{2}}^{2}/4$, i.e., a two-arm interferometer with the unequal weights $1$ and $\abs{R}^{2}=1-P_{01}$.
Expanding the modulus with $R^{2}=\abs{R}^{2}e^{2i\chi}$ separates the two,
\begin{equation}\label{eq:P_ee_chi}
    \begin{split}
        P_{ee}(\tau)&=\frac{\left(1-\abs{R}^{2}\right)^{2}}{4}+\abs{R}^{2}\sin^{2}\chi\\
        &=\frac{P_{01}^{2}}{4}+\left(1-P_{01}\right)\sin^{2}\chi,
    \end{split}
\end{equation}
and likewise $P_{oo}(\tau)=P_{01}^{2}/4+\left(1-P_{01}\right)\cos^{2}\chi$, both exact for ideal pulses and with the entire fringe in the phase $\chi$.
Their sum leaves the leakage probability after the sequence,
\begin{equation}\label{eq:P_leak_ramsey}
    \begin{split}
        P_{\text{leak}}(\tau_{\text{wait}})&=1-P_{ee}-P_{oo}=\frac{1-\left[1-P_{01}(\tau_{\text{wait}})\right]^{2}}{2}\\
        &=P_{01}-\frac{P_{01}^{2}}{2}\\
        &\approx\frac{\abs{E_-}}{\Delta}\sin^{2}\left(\frac{\left(2\Delta+\abs{E_-}\right)\tau_{\text{wait}}}{2\hbar}\right),
    \end{split}
\end{equation}
which is the weight of $\ket{\psi_{\text{leak}}}$ in Eq.~\eqref{eq:psi3_ramsey} and does not depend on $\chi$; the last line is the first order in $\abs{E_-}/\Delta$.
Equation~\eqref{eq:P_leak_ramsey} is Eq.~\eqref{eq:P_leak_ramsey_main} of Sec.~\ref{sec:ramsey}.
Splitting the bracket of Eq.~\eqref{eq:R_ramsey} into counter-rotating parts, $\cos\phi+ik\sin\phi=\frac{1+k}{2}e^{i\phi}+\frac{1-k}{2}e^{-i\phi}$, gives
\begin{equation}\label{eq:chi_expand}
    \begin{split}
        \chi&=\underbrace{-\frac{\Delta\tau}{\hbar}+\phi}_{\abs{E_-}\tau/2\hbar}+\arg\left(1+\frac{\abs{E_-}}{4\Delta+\abs{E_-}}e^{-2i\phi}\right)\\
        &=\frac{\abs{E_-}\tau}{2\hbar}-\frac{\abs{E_-}}{4\Delta+\abs{E_-}}\sin\frac{\left(2\Delta+\abs{E_-}\right)\tau}{\hbar}+\mathcal{O}\left(\frac{E_-^2}{\Delta^2}\right).
    \end{split}
\end{equation}
The factor $e^{-i\Delta\tau/\hbar}$ in Eq.~\eqref{eq:R_ramsey} cancels the precession of the dark branch, so each chain contributes the difference frequency $\abs{E_-}/2$, and the secular part of the two-chain phase $2\chi$, i.e., the part growing linearly in $\tau_{\text{wait}}$, is $\abs{E_-}\tau_{\text{wait}}/\hbar$; the remainder is non-secular, i.e., a bounded oscillation of $\chi$ at the frequency $\left(2\Delta+\abs{E_-}\right)/\hbar$ with amplitude $\abs{E_-}/4\Delta$ that does not accumulate with $\tau_{\text{wait}}$.
Dropping this non-secular term in Eq.~\eqref{eq:P_ee_chi}, the return probability after the sequence is
\begin{equation}\label{eq:P_ee_ramsey}
    P_{ee}(\tau_{\text{wait}})\approx\left[1-P_{01}(\tau_{\text{wait}})\right]\sin^2\left(\frac{E_-\tau_{\text{wait}}}{2\hbar}\right)+\frac{P_{01}^2}{4},
\end{equation}
which is Eq.~\eqref{eq:P_ee_ramsey_main} of Sec.~\ref{sec:ramsey}, where the floor $P_{01}^2/4=\left[\left(1-\abs{R}^2\right)/2\right]^2$ is the arm imbalance rather than a return of the leaked population, and Eq.~\eqref{eq:P_ee_ramsey} reduces to $\sin^2\left(E_-\tau_{\text{wait}}/2\hbar\right)$ for $\abs{E_-}\ll\Delta$.
Similarly, for $E_-<0$, the roles of the even and odd sectors are exchanged (Appendix~\ref{app:ramsey_N3}), so that $\ket{e}=\ket{00}$ leaks to $\ket{11}$ with the same probability Eq.~\eqref{eq:P01} while $\ket{o}$ stays dark; the $\ket{ee}$ branch then carries the factor $R^2$ of Eq.~\eqref{eq:R_ramsey} instead of $\ket{oo}$, which leaves Eqs.~\eqref{eq:P_ee_chi}--\eqref{eq:P_ee_ramsey} unchanged.
In contrast to the Rabi leakage, which is second order in $\Gamma/\Delta$, the Ramsey leakage is first order in $\abs{E_-}/\Delta$ and fixed by the target splitting itself, and it oscillates at half the frequency of the Rabi leakage in Fig.~\ref{fig:pristine_rabi}(c,e), because the wait excites one bulk quasiparticle across the gap $2\Delta$ rather than a pair costing $4\Delta$.
This is not an additional pathology but the operational consequence of removing every first-order splitting channel at the genuine three-site sweet spot.
Leaving the sweet spot for the wait also gives up the protection of Eq.~\eqref{eq:epsilon_N3}: once $E_-\neq0$, local fluctuations shift $E_-$ at first order instead of second.

\section{Junction coupling $\Gamma$ for the experimental parameter set}\label{app:Gamma_exp}
In this section, we derive the mean and the standard deviation of $\Gamma$ in the experimental column of Table~\ref{tab:parameters}.
In Ref.~\cite{zatelli2026majorana}, the two chains are not tunnel coupled directly as in Eq.~\eqref{eq:Htunn}, but through an additional quantum dot $C$ between $L_N$ and $R_1$,
\begin{equation}\label{eq:Htunn_C}
    \mathcal{H}_{\text{tunn}}^{C}=\mu_{C}n_{C}+t_{LC}c_{C}^{\dagger}c_{L_N}+t_{CR}c_{R_1}^{\dagger}c_{C}+\hc,
\end{equation}
with $t_{LC}=t_{CR}=7.5\ueV$ and $\mu_{C}=130\ueV$ (Table~1 of Ref.~\cite{zatelli2026majorana}).
Since $\mu_{C}\gg t_{LC},t_{CR}$, the dot $C$ remains empty and is integrated out at second order in $t_{LC}/\mu_{C}$ and $t_{CR}/\mu_{C}$, which reduces Eq.~\eqref{eq:Htunn_C} to the direct tunneling Eq.~\eqref{eq:Htunn} with the effective amplitude
\begin{equation}\label{eq:Gamma_eff}
    \Gamma=\frac{t_{LC}t_{CR}}{\mu_{C}}=0.433\ueV,
\end{equation}
up to a sign, which is immaterial [cf. the redefinition of $\ket{oo}$ in Appendix~\ref{app:effective_model}], and the on-site shifts $-t_{LC}^{2}/\mu_{C}$ and $-t_{CR}^{2}/\mu_{C}$ of $\mu_{L_N}$ and $\mu_{R_1}$, which are absorbed into the sweet-spot condition Eq.~\eqref{eq:sweetspot}.
The disorder in $\Gamma$ follows from that of $t_{LC}$, $t_{CR}$, and $\mu_{C}$, with $\sigma_{t_{LC}}=\sigma_{t_{CR}}=0.1\ueV$ and $\sigma_{\mu_{C}}=4.5\ueV$~\cite{zatelli2026majorana}, which to leading order propagates as
\begin{equation}\label{eq:sigma_Gamma}
    \sigma_{\Gamma}=\Gamma\sqrt{\left(\frac{\sigma_{t_{LC}}}{t_{LC}}\right)^{2}+\left(\frac{\sigma_{t_{CR}}}{t_{CR}}\right)^{2}+\left(\frac{\sigma_{\mu_{C}}}{\mu_{C}}\right)^{2}}=0.017\ueV.
\end{equation}
Equations~\eqref{eq:Gamma_eff} and \eqref{eq:sigma_Gamma} are the $\Gamma$ entry of the experimental column in Table~\ref{tab:parameters}.

\section{Dephasing envelope of a phase quadratic in the disorder}\label{app:geff_average}
In this section, we derive the algebraic envelopes Eq.~\eqref{eq:rabi_perp} of Sec.~\ref{sec:renormalization_tunneling} by averaging a phase that is quadratic in the Gaussian disorder; Eq.~\eqref{eq:rabi_eps} of Sec.~\ref{sec:interchain_splitting} is the same average with a single variable.

Let $\bm{Z}=(Z_1,\dots,Z_n)$ be independent standard normal variables and $\varphi=\theta\,\bm{Z}^\intercal A\bm{Z}$ with $A$ real symmetric.
An orthogonal rotation to the eigenbasis of $A$ preserves the distribution of $\bm{Z}$, so $\varphi=\theta\sum_{k=1}^{n}\lambda_kZ_k^2$ with $\lambda_k$ the eigenvalues of $A$.
Since $\expval{e^{i\alpha Z^2}}=(1-2i\alpha)^{-1/2}$ for $Z\sim\mathcal{N}(0,1)$, i.e., the characteristic function of a chi-squared variable with one degree of freedom, the average factorizes,
\begin{equation}\label{eq:quad_char}
    \begin{split}
        \expval{e^{i\varphi}}&=\prod_{k=1}^{n}\left(1-2i\lambda_k\theta\right)^{-1/2},\\
        \abs{\expval{e^{i\varphi}}}&=\prod_{k=1}^{n}\left(1+4\lambda_k^2\theta^2\right)^{-1/4}.
    \end{split}
\end{equation}
The phase of Eq.~\eqref{eq:quad_char}, $\frac{1}{2}\sum_k\arctan\left(2\lambda_k\theta\right)\approx\expval{\varphi}$, renormalizes the oscillation frequency, and the modulus is the damping envelope.
The envelope is even in each $\lambda_k$, so the signs of the eigenvalues are irrelvant.
At short times, $\abs{\expval{e^{i\varphi}}}\approx1-\theta^2\sum_k\lambda_k^2=1-\operatorname{Var}(\varphi)/2$ depends only on the variance, whereas the tail $\abs{\expval{e^{i\varphi}}}\propto\theta^{-n/2}$ is set by the rank $n$ of $A$.
For $\abs{\lambda_k}=\lambda$, the $1/e$ time of Eq.~\eqref{eq:quad_char} is
\begin{equation}\label{eq:quad_T2}
    \theta_{1/e}=\frac{\sqrt{e^{4/n}-1}}{2\lambda},
\end{equation}
which decreases with $n$ at fixed $\lambda$.

For the intrachain splitting channel, $\varphi=E_-^2t/2\Gamma\hbar$ with $E_-\approx\sigma_{E_-}Z$, so $n=1$, $\lambda=1$, and $\theta=\sigma_{E_-}^2t/2\Gamma\hbar$; Eqs.~\eqref{eq:quad_char} and \eqref{eq:quad_T2} give the envelope $\left[1+\left(\sigma_{E_-}^2t/\Gamma\hbar\right)^2\right]^{-1/4}$ and Eq.~\eqref{eq:rabi_eps}.

For the renormalized tunneling channel, $\varphi=\delta\Gamma_{\text{eff}}t/\hbar$ and $\theta=x/2$ with $x=\Gamma\sigma_\mu^2t/[(2\Delta)^2\hbar]$, up to an overall sign.
For $N=3$ and $\sigma_\nu\ll\sigma_\mu$, Eq.~\eqref{eq:gamma_eff_N3} gives $A=\mathds{1}$ in the variables $(\mu_{L_3},\mu_{R_1})/\sigma_\mu$, so $n=2$, and the omitted $\nu_{a_i}$ terms contribute four factors to Eq.~\eqref{eq:quad_char} with $\abs{\lambda_k}=\sigma_\nu^2/\sigma_\mu^2$, which deviate from unity only for $x\gtrsim\sigma_\mu^2/\sigma_\nu^2\gg1$.
For $N=2$, the quadratic form of Eq.~\eqref{eq:gamma_eff_N2} in the variables $(\mu_{L_2},\mu_{R_1},\mu_{L_1},\mu_{R_2})/\sigma_\mu$ has the eigenvalues $(1,1,1,-1)$, the cross term being diagonalized by $2\mu_{L_1}\mu_{R_2}=\frac{1}{2}\left[\left(\mu_{L_1}+\mu_{R_2}\right)^2-\left(\mu_{L_1}-\mu_{R_2}\right)^2\right]$, so $n=4$.
Equation~\eqref{eq:quad_char} with $\lambda=1$ gives the two envelopes $(1+x^2)^{-n/4}$ of Eq.~\eqref{eq:rabi_perp}, Eq.~\eqref{eq:quad_T2} gives Eq.~\eqref{eq:T2_geff}, and the common mean shift $\expval{\delta\Gamma_{\text{eff}}}/\Gamma=-\sigma_\mu^2/(2\Delta)^2$ renormalizes the Rabi frequency for both $N$.
Although the variance of $\delta\Gamma_{\text{eff}}^{(2)}$ is only twice that of $\delta\Gamma_{\text{eff}}^{(3)}$, the ratio of the $1/e$ times is $\sqrt{e^2-1}/\sqrt{e-1}=\sqrt{e+1}$ rather than $\sqrt2$, because the $1/e$ time is set by the tail of Eq.~\eqref{eq:quad_char} rather than by its short-time expansion.

\end{document}